 \documentclass[12pt]{article}
 \newcommand{\be}{\begin{equation}}
 \newcommand{\ee}{\end{equation}}
 \newcommand{\ba}{\begin{eqnarray}}
 \newcommand{\ea}{\end{eqnarray}}
 \newcommand{\inc}{{\it i}}
 \newcommand{\bra}{\langle}
 \newcommand{\ket}{\rangle}
 \newcommand{\efbold}{\mbox{{\boldmath $\vec f$}}}

 \newcommand{\mubold}{\mbox{{\boldmath $\vec \mu$}}}
 \newcommand{\omegabold}{\mbox{{\boldmath $\vec \omega$}}}

 \newcommand{\wbold}{\mbox{{\boldmath $\vec w$}}}
 \newcommand{\dotmubold}{\stackrel{\textbf{.}}{{\bf{
                      \mbox{\boldmath${\boldmath\vec{\boldmath{\mu}}}$}}}}}

\begin{document}
 \title{
 {\Large{\textbf{Long-term evolution of orbits about a
 precessing\footnote{~By ``precession," in its
 most general sense, we mean any
 change of the direction of the spin axis of the planet --
 from its long-term variations down to
 nutations down to the Chandler wobble and polar wander.} oblate planet.\\
 2. The case of variable precession.}
            }}}
 \author{
 {\Large{Michael Efroimsky}}\\
 {\small{US ~Naval ~Observatory, ~Washington ~DC ~20392 ~USA, ~~~e-mail: me @
 usno.navy.mil~}}\\
% ~\\
% {\Large{Victor Slabinski}}\\
% {\small{US Naval Observatory, Washington DC 20392 USA, ~~e-mail:
% slabiv @ tos.usno.navy.mil~}}\\
% ~\\
% and\\
% ~\\
% {\Large{Marc Murison}}\\
% {\small{US Naval Observatory, Washington DC 20392 USA}}\\
% {\small{e-mail: murison @ arnold.usno.navy.mil~}}
 }

 \maketitle

 \begin{abstract}
 We continue the study undertaken in Efroimsky (2005a) where we
 explored the influence of spin-axis variations of an oblate planet on satellite orbits.
 Near-equatorial satellites had long been believed to keep up with the oblate
 primary's equator in the cause of its spin-axis variations. As demonstrated by Efroimsky
 \& Goldreich (2004), this opinion had stemmed from an inexact interpretation of a correct
 result by Goldreich (1965). Though Goldreich (1965) mentioned that his result
 (preservation of the initial inclination, up to small oscillations about the moving
 equatorial plane) was obtained for {{\emph{non-osculating}}} inclination, his admonition
 has been persistently ignored for forty years.

 It was explained in Efroimsky \& Goldreich (2004) that the equator precession influences
 the osculating inclination of a satellite orbit already in the first order over the
 perturbation caused by a transition from an inertial to an equatorial coordinate system.
 It was later shown in Efroimsky (2005a) that the {\it{secular part}} of the inclination
 is affected only in the second order. This fact, anticipated by Goldreich (1965), remains
 valid for a constant rate of the precession. It turns out that non-uniform variations of
 the planetary spin state generate changes in the osculating elements, that are linear in
 $\;|\dotmubold |\;$ (where $\;\mubold\;$ is the planetary equator's total precession
 rate, rate that includes the equinoctial precession, nutation, the Chandler wobble, and
 the polar wander).

 % The effect is enforced by (though not reduced to) commensurabilities between the
 % satellite orbital frequency and the planetary short-period nutations. Another relevant
 % physical effect is the precession of the planet's orbit about the Sun. It leads to
 % variations in the Solar torque acting on the planet; the torque variations cause
 % changes in the planet axis' equinoctial precession; these changes, in their turn,
 % entail variations of orbits of the planet's satellite. This effect is extremely weak
 % and accumulates over very long time scales.

 We work out a formalism which will help us to determine if these factors cause a drift of
 a satellite orbit away from the evolving planetary equator.\\
\\
\end{abstract}

\section{The scope of the project}

\subsection{The motivation}

% Since the obliquity of a planet strongly influences the latitudinal
% distribution of its yearly insolation, the obliquity history is
% crucial to the climate-evolution studies and, thereby, to evaluation
% of the chances for life emergence -- an issue especially relevant to
% Mars. The methods used in obliquity-history research
 Calculation of the obliquity of a planet
% were pioneered by Sharaf \& Budnikova (1969a, b), Vernekar (1972),
 (Ward 1973;
% by Ward (1973, 1974, 1979, 1982) who studied the spin-orbit
% resonance between the precession rate of Mars' rotation axis and
% one of the frequencies in its orbital precession. Ward's analysis
% has demonstrated that Mars' obliquity had been varying with a $10^o$
% amplitude over time scales of $10^{5}\,-\,10^6$ years.
 Laskar \& Robutel 1993; Touma \& Wisdom 1994) are always
obtained within a simplified model based on representation of the
planet by a symmetrical rigid rotator, with no internal structure or
dissipative phenomena taken into consideration. This model yields,
via the Colombo (1966) equation, the history of the planet axis'
inclination in an inertial frame. Thence the evolution of the
obliquity
% (i.e., of the inclination on the precessing ecliptic)
can be found. The Colombo (1966) equation was derived for a rigid
planet in the principal rotation state. These assumptions raise
questions when it comes to real physics. First, a planet is
deformable and, thereby, is subject to solar tides. It also tends to
yield its shape to the instantaneous axis of rotation. (This
phenomenon is always acknowledged in regard to the Chandler wobble,
but never in regard to the equinoctial precession.) Second, a forced
rotator is never in a principal spin state, and its angular-velocity
vector is always slightly off its angular-momentum vector. These
three phenomena influence the equinoctial precession and, through
it, the obliquity variations. On the one hand, these phenomena are
feeble; on the other hand, we are interested in their accumulation
over the longest time scales, and therefore we are unsure of the
outcome. Last, and by no means least, the Colombo description of the
equinoctial precession ignores the possibility of planetary
catastrophes that might have altered the planet's spin mode.

 % On these grounds,
 It would be good to develop a model-independent check of whether
 the planet could have maintained, through its entire past, the same equinoctial
 precession as it has today. Such a check is
 % immediately
 offered by Mars' two satellites.
 % Their existence in nearly equatorial orbits poses the question: how have their orbital
 % planes evolved in the course of Mars' equinoctial precession?
 The present proximity
 of both moons to the Martian equatorial plane is hardly a mere coincidence.
 Hence, the question becomes: could Mars have maintained equinoctial precession,
 predicted by the Colombo model, through its entire history without pushing an initially
 near-equatorial satellite too far away from the equatorial plane?

% The dependence of the obliquity history upon the underlying model
% of equinoctial precession necessitates a quest for a
% model-independent test, a check that would enable one to approach
% the problem from a totally different direction.

\subsection{The objective}

 If it turns out that the equinoctial precession, predicted by the Colombo (1966) model,
 does not drive the satellites away from the equator, or drives them away at a very slow
 rate, then this will become an independent confirmation of this model's
 applicability. If, however, it turns out that the predicted precession of the spin axis
 leads to considerable variations in the satellite inclination relative to the equator of
 date, this will mean that the Colombo model should be further improved
% by taking into consideration dissipation or/and
% the internal structure of Mars.
 or/and that a planetary catastrophe may have altered Mars' spin state.

 According to Goldreich (1965) and Kinoshita (1993)
% , Waz (1999),
 the inclination of a near-equatorial satellite only oscillates about
 its initial value, provided the equinoctial precession is uniform. However, even within
 the simple Colombo model, the equinoctial precession is variable.
 Besides, in these works {\emph{non-osculating}} elements were used, circumstance noticed
 by Goldreich (1965) but missed by many authors who employed and furthered his result.

 Whenever the disturbance depends upon velocities
 (like a transition from inertial axes to ones
 co-precessing with the planet), a mere amendment of the disturbing function
 % (by adding to it the Hamiltonian variation taken with the opposite sign)
 makes the planetary equations
 render not the osculating but the so-called contact orbital elements whose
 physical interpretation is nontrivial (Efroimsky \& Goldreich 2004). To furnish
 osculating elements, the equations should be enriched with
 extra terms, some of which will not be additions to the disturbing
 function.\footnote{~These terms will complicate both the Lagrange- and
 Delaunay-type equations. The Delaunay equations will no longer be
 Hamiltonian. This parallels a predicament with the Andoyer elements
 used in the theory of rigid-body rotation with angular-velocity-dependent
 perturbations (Efroimsky 2007; Gurfil, Elipe, Tangren, \& Efroimsky 2007)} Some of them
 will be of the \emph{first order} in the velocity-dependent perturbation, others
 of the second. For {\it{uniform}} precession, the first-order extra terms average
 out, except for a term showing up in the equation for $\,dM_o/dt$
 (Efroimsky 2005a), as predicted by Goldreich forty years ago.
 Thus, if we address not the elements {\emph{per se}} but
 their {\emph{secular parts}},
 % and for not too long time intervals,
 Goldreich's result obtained for the contact elements stays
 also for the osculating ones: the orbit will oscillate about the uniformly
 moving equator but will not shift away from it.

As demonstrated in Efroimsky (2005a), under {\it{variable}}
precession of the spin axis, the secular parts of these
precession-caused first-order terms are of the first order.
Accordingly, the secular parts of the osculating elements may differ
from those of the contact ones already in this order.
% In particular, what was valid for the contact inclination will not
% necessarily be valid for the osculating inclination: the orientation
% of the satellite orbit relative to the moving equatorial plane will
% no longer be obliged to be limited to small oscillations only.

To understand if Mars could have kept through its entire past the
same equinoctial precession, we need to determine if the satellite
orbits might have shifted away from the equator in the cause of
nonuniform precession. To see how the secular parts of the
osculating elements evolve, we shall build the required mathematical
formalism
 % While an impeccably rigorous analysis will demand numerical integration
 % of the complete planetary equations,
based on the averaged equations.

\subsection{The means}

The motion of a satellite about a precessing oblate planet is most
naturally described with orbital elements defined in a
co-precessing (equatorial) coordinate system. It is also
convenient to choose the elements to be osculating. The physical
interpretation of such orbital variables will be most
straightforward.

\subsubsection{\textbf{Exact planetary equations}}

The above defined setting is the two-body problem disturbed by two
perturbations -- the gravitational pull of the equatorial bulge
and the transition to a non-inertial frame of reference associated
with the precessing planetary equator. Together, they generate the
following variation of the Hamiltonian (Efroimsky \& Goldreich
2003, Efroimsky 2005a,b):
 \ba
 \Delta {\cal H}^{(osc)}\;=\;-\,\left[\;R_{oblate}(\nu)\,+\,
 \mubold\cdot(\efbold\times{\bf{\vec g}})\;+\;
 (\mubold\times\efbold)\cdot(\mubold\times\efbold)\;\right]\;\;\;,
 \label{1}
 \ea
where the oblateness-caused disturbing potential is
  \ba
  R_{oblate}(\nu )\;=\;\frac{G\,m\;J_2}{2}\;\frac{\rho_e^2}{r^3}\;\left[\;1\;-
  \;3\;\sin^2 \inc\;\sin^2(\omega\;+\;\nu) \;\right]\;\;,
  \label{2}
  \ea
$\rho_e\;$ being the mean equatorial radius of the planet, and
$\;\nu\;$ standing for the true anomaly. The vector
 \ba
 \mubold\;=\;\mu_1\,{\bf{\hat x}}\;+\;\mu_2\;
 {\bf{\hat y}}\;+\;\mu_3\;{\bf{\hat
 z}}\;=\;{\bf{\hat x}}\,\frac{d I_p}{dt}\;+\;
 {\bf{\hat y}}\,\frac{dh_p}{dt}\,\sin I_p\;+\;{\bf{\hat
 z}}\,\frac{dh_p}{dt}\,\cos I_p\;\;\;,
 \label{3}
 \ea
is the precession rate of the planetary spin axis. (In the
astronomical literature it is sometimes referred to as the
rotational vector of the equator.) Angles $\;I_p\;$ and $\;h_p\;$
are the inclination and the longitude of the node of the planetary
equator of date relative to that of epoch; unit vectors
$\;{\bf{\hat x}}\,,\;{\bf{\hat y}}\,,\;{\bf{\hat z}}\;$ denote a
coordinate system fixed on the moving equatorial plane of date,
$\;{\bf{\hat z}}\;$ being orthogonal to the equator-of-date plane,
and $\;{\bf{\hat x}}\;$ pointing toward the ascending node of the
equator of date relative to the one of epoch. For details of
calculation of $\;I_p\;$ and $\;h_p\;$ see subsection 2.2.2 and
the Appendix.

Notations $\;\efbold\;$ and $\;\bf\vec g\;$ stand for two
auxiliary vector functions which play an important role in the
theory. These are the implicit functional dependencies of the
unperturbed (two-body) position and velocity upon time and six
orbital elements. These dependencies emerge as a solution
  \begin{eqnarray}
  \nonumber
  {\bf \vec r }\;=\;{\efbold} \left(C_1, ... , C_6, \,t
\right)\,\;\;\;\,~~~~~~~~~~~~~~~~~~\\
 \label{4}\\
 \nonumber
   {\bf \vec v }\;=\;{\bf\vec g} \left(C_1, ... , C_6, \,t
\right)\;\;\;,\,\;\;\;\,\;\;{\bf\vec g}\;\equiv\;\frac{\partial
\efbold}{\partial t}
  \end{eqnarray}
to the reduced two-body problem
  \begin{equation}
  {\bf \ddot{\vec
r}}\;+\;\frac{G\,m}{r^2}\;\frac{{\bf\vec r }}{r}\;=\;0\;\;\;\,
 \label{5}
  \end{equation}
and define, geometrically, a Keplerian ellipse or hyperbola
parameterised with some set of six independent orbital elements
$\;C_i\;$ which are constants in the absence of disturbances.
Under perturbation, these elements are endowed with time
dependence.

This way, our Hamiltonian variation $\;\Delta
{\cal{H}}^{(osc)}\;$, too, becomes, through composition, a
function of time and the same six orbital elements used in
(\ref{4}) (these could be the Keplerian or Delaunay or Poincare or
Jacobi elements). The Hamiltonian variation is equipped with
superscript $\,``{(osc)}"\,$ in order to emphasise that this is
the form taken by the Hamiltonian expressed as a function of
{\underline{osculating}} orbital elements. This clause, seemingly
trivial and therefore redundant, turns out to be crucially
important. As pointed out in Efroimsky \& Goldreich (2004) and
explained in great detail in Efroimsky (2005a), a naive
development of the planetary equations in precessing frames leads
to a Hamiltonian variation different from (\ref{1}); but that
Hamiltonian variation tacitly turns out to be a function of
non-osculating orbital elements. This tacit loss of osculation in
problems with velocity-dependent perturbations is an old pitfall
in orbit calculations. Though some 40 years ago Goldreich (1965)
warned of these difficulties, the issue has until recently been
ignored in the literature. This has led to a whole sequence of
erroneous results. (As we already mentioned above, a very similar
situation emerges in the rigid-body attitude dynamics.)

For some general-type parameterisation of the instantaneous conics
through six orbital variables $\;C_1\,,\;.\;.\;.\;,\,C_6\;$, the
variation-of-parameters equations will read
 \begin{eqnarray}
 \nonumber [C_n\;C_i]\;\frac{dC_i}{dt}\;=\;-\;
 \frac{\partial\,\Delta {\cal H}^{(osc)}}{\partial
 C_n}\,~~~~~~~~~~~~~~~~~~~~~~~~~
 \nonumber \\ &&
 \label{6}\\
 \nonumber +\;\mubold\cdot \left(\frac{\partial{\efbold}}{\partial
 C_n}\times {\bf{\vec g}}\;-\;{\efbold}\times \frac{\partial{\vec{\bf
 g}}}{\partial C_n}\right)\;-\; {\bf{\dot{\mubold
 }}}\cdot\left(\efbold\times \frac{\partial \efbold }{\partial
 C_n}\right)\;-\;\left(\mubold\times\efbold\right) \;\frac{\partial
 }{\partial C_n}\left(\mubold\times\efbold\right) \;\;\;.
 \end{eqnarray}
provided these conics are chosen to be always tangent to the
trajectory, i.e., provided the parameters are chosen to be
osculating\footnote{~Had we simply amended the Hamiltonian by the
above variation $\;\Delta {\cal H}^{(osc)}\;$, without inserting
the extra $\;\mubold $-dependent terms into (\ref{6}), such
equations would yield non-osculating elements, ones parametrising
a family of non-tangent conics. This would happen because the
Hamiltonian perturbation depends not only upon positions but also
upon the canonical momenta. Another way of getting into this
hidden trap is to start with the Cartesian or spherical
coordinates and momenta, and to perform the Hamilton-Jacobi
operation. The resulting variables $\;C_j\;$ will then come out
canonical and will be the well-known Delaunay elements. In case
the Hamiltonian perturbation depends upon the momenta, these
Delaunay elements will be non-osculating, i.e., will parameterise
a sequence of instantaneous conics non-tangent to the physical
orbit. (Efroimsky \& Goldreich 2003; Efroimsky 2005a)} (Efroimsky
\& Goldreich 2004, Efroimsky 2005a).

A more convenient representation of the above equation will be
achieved if one includes the
$\;\;-\,\left(\mubold\times\efbold\right) \;{\partial
 \left(\mubold\times\efbold\right) }/{\partial C_n} \;$ term
 in the Hamiltonian:
  \begin{eqnarray}
 % \nonumber
 [C_n\;C_i]\;\frac{dC_i}{dt}\;=\;-\;
 \frac{\partial\;``\Delta {\cal H}"}{\partial C_n}~
 % ~~~~~~~~~~~~~~~
 % \nonumber \\ &&
 % \label{40}\\
 % \nonumber
 +\;\mubold\cdot
 \left(\frac{\partial{\efbold}}{\partial C_n}\times
 {\bf{\vec g}}\;-\;{\efbold}\times
 \frac{\partial{\vec{\bf g}}}{\partial C_n}\right)\;-\;
 {\bf{\dot{\mubold }}}\cdot\left(\efbold\times
 \frac{\partial \efbold }{\partial C_n}\right)\;\;\;,\;\;\;\;
 \label{7}
 \end{eqnarray}
the amended ``Hamiltonian" being defined through
 \begin{equation}
 ``\Delta {\cal H}"\;=\;-\,\left[\;R_{oblate}(\nu)\,+\,
 \mubold\cdot(\efbold\times{\bf{\vec g}})\;+\;\frac{1}{2}\;
 (\mubold\times\efbold)\cdot(\mubold\times\efbold)\;\right]\;\;\;
 \label{8}
 \end{equation}
Here the quotation marks are necessary to emphasise that
$\;``\Delta{\cal H}"\;$ is not the real Hamiltonian variation but
merely a convenient mathematical entity. Under this convention,
and under the assumption that the parameterisation is implemented
through the Kepler elements, (\ref{7}) yields the following system
of Lagrange-type planetary equations:
  ~\\
 \ba
 \frac{da}{dt}\,=\,
 \frac{2}{n a}\,\left[\,\;\frac{\partial\left(\,-\,``\Delta {\cal H}"\right)
 }{\partial M_o}\,-\,
 {\bf{\dot{\mubold }}}\cdot\left(\efbold\times
 \frac{\partial \efbold }{\partial
 M_o}\right)\;\right]\;\;,\;\;\;~~~~~~~~~~~~~~~~~~~~~~~~~~~~~~~~~~~~~~~~~~~
 \label{9}
  \ea
  ~\\
 ~\\
  \begin{eqnarray}
  \nonumber
 \frac{de}{dt}\,=\,\frac{1-e^2}{n\,a^2\,e}\;\;\left[\;\frac{\partial
 \left(\,-\,``\Delta {\cal H}" \right)  }{\partial M_o}
 \,-\,
 {\bf{\dot{\mubold }}}\cdot\left(\efbold\times
 \frac{\partial \efbold }{\partial M_o}\right)
 \;\right]~~\,~~~~~~~~~~~~~~~~~~~~~~~~~~~~~~~~~~~~~~~~~\\
 \nonumber\\
 \label{10}\\
 \nonumber\\
 \nonumber
 -\;
 \frac{(1\,-\,e^2)^{1/2}}{n\,a^2\,e} \;\left[\;\frac{\partial
 \left(\,-\,``\Delta {\cal H}" \right) }{\partial \omega} \,~ +\, \mubold\cdot
 \left(\frac{\partial{\efbold}}{\partial \omega}\times
 {\bf{\vec g}}\,-\,{\efbold}\times
 \frac{\partial{\vec{\bf g}}}{\partial \omega}\right)\,-\,
 {\bf{\dot{\mubold }}}\cdot\left(\efbold\times
 \frac{\partial \efbold }{\partial \omega}\right)
 \;\right]~~,~~~
  \end{eqnarray}
 ~\\
 ~\\
  \begin{eqnarray}
 \nonumber
 \frac{d\omega}{dt}\,=~
 \frac{\;-\,\cos \inc
}{n a^2(1-e^2)^{1/2}\, \sin \inc }\left[ \frac{\partial
\left(\,-\,``\Delta {\cal H}" \right) }{\partial \inc } \,~ + ~
\mubold\cdot
 \left(\frac{\partial{\efbold}}{\partial \inc}\times
 {\bf{\vec g}}\,-\,{\efbold}\times
 \frac{\partial{\vec{\bf g}}}{\partial \inc}\right)\,-\,
 {\bf{\dot{\mubold }}}\cdot\left(\efbold\times
 \frac{\partial \efbold }{\partial \inc}\right)\,
  \right]~~~~\\
 \nonumber\\
 \label{11}\\
 \nonumber\\
 \nonumber
 +\,~\frac{(1-e^2)^{1/2}}{n\,a^2\,e}\;
 \left[\;\frac{\partial \left(\,-\,``\Delta {\cal H}" \right) }{\partial e}\,
+ ~ \mubold\cdot
 \left(\frac{\partial{\efbold}}{\partial e}\times
 {\bf{\vec g}}\,-\,{\efbold}\times
 \frac{\partial{\vec{\bf g}}}{\partial e}\right)\,-\,
 {\bf{\dot{\mubold }}}\cdot\left(\efbold\times
 \frac{\partial \efbold }{\partial e}\right)
 \;\right]\;\;,~~~~~~~~~~~
  \end{eqnarray}
 ~\\
 ~\\
 ~\\
 \ba
 \nonumber
 \frac{d \inc
 }{dt}\;=~ \frac{\cos \inc}{n a^2\,(1 - e^2)^{1/2} \sin
 \inc}
 \left[\frac{\partial \left(\,-\,``\Delta {\cal H}" \right)}{\partial
 \omega}\;+ \; \mubold\cdot
 \left(\frac{\partial{\efbold}}{\partial \omega}\times
 {\bf{\vec g}}\,-\,{\efbold}\times
 \frac{\partial{\vec{\bf g}}}{\partial \omega}\right)\,-\,
 {\bf{\dot{\mubold }}}\cdot\left(\efbold\times
 \frac{\partial \efbold }{\partial \omega}\right)
 \,\right]\,-\;\;\;\\
 \nonumber\\
 \label{12}\\
 \nonumber\\
 \nonumber
 \frac{1}{na^2\,(1-e^2)^{1/2}\,\sin \inc
 }\,\left[\,\frac{\partial \left(\,-\,``\Delta {\cal H}" \right) }{\partial
 \Omega}\;+ \; \mubold\cdot
 \left(\frac{\partial{\efbold}}{\partial \Omega}\times
 {\bf{\vec g}}\,-\,{\efbold}\times
 \frac{\partial{\vec{\bf g}}}{\partial \Omega}\right)\,-\,
 {\bf{\dot{\mubold }}}\cdot\left(\efbold\times
 \frac{\partial \efbold }{\partial \Omega}\right)
 \,\right]\;{,}\;\;\;
  \ea
 ~\\
 ~\\
  \ba
\frac{d\Omega}{dt}\,=~\frac{1}{n a^2\,(1-e^2)^{1/2}\,\sin \inc
}\,\left[\,\frac{\partial \left(\,-\,``\Delta {\cal H}" \,\right)
}{\partial \inc }\,~ + ~\mubold\cdot
 \left(\frac{\partial{\efbold}}{\partial \inc}\times
 {\bf{\vec g}}\,-\,{\efbold}\times
 \frac{\partial{\vec{\bf g}}}{\partial \inc}\right)\,-\,
 {\bf{\dot{\mubold }}}\cdot\left(\efbold\times
 \frac{\partial \efbold }{\partial \inc}\right)\,
 \right]\;,\;\;\;
 \label{13}
  \ea
 ~\\
 ~\\
  \begin{eqnarray}
  \nonumber
\frac{dM_o}{dt}\,=\,\;-\,\;\frac{1\,-\,e^2}{n\,a^2\,e}\,\;
\left[\;\frac{\partial \left(\,-\,``\Delta {\cal H}" \right)
 }{\partial e }~ + ~
 \mubold\cdot
 \left(\frac{\partial{\efbold}}{\partial e}\times
 {\bf{\vec g}}\,-\,{\efbold}\times
 \frac{\partial{\vec{\bf g}}}{\partial e}\right)\,-\,
 {\bf{\dot{\mubold }}}\cdot\left(\efbold\times
 \frac{\partial \efbold }{\partial e}\right)
 \;\right] ~~~~~\\
 \nonumber\\
 \label{14}\\
 \nonumber\\
 \nonumber
 -\;~\frac{2}{n\,a}\,\left[\;\frac{\partial \left(\,-\,``\Delta {\cal H}" \right)
 }{\partial a }~ + ~ \mubold\cdot
 \left(\frac{\partial{\efbold}}{\partial a}\times
 {\bf{\vec g}}\,-\,{\efbold}\times
 \frac{\partial{\vec{\bf g}}}{\partial a}\right)\,-\,
 {\bf{\dot{\mubold }}}\cdot\left(\efbold\times
 \frac{\partial \efbold }{\partial a}\right)
 \;\right]
 \;\;,~~~~~
  \end{eqnarray}
  ~\\
  where terms $\;\mubold\cdot\left(\;(\partial \efbold/\partial M_o)\times {\bf\vec g}\,-\,
  (\partial {\bf \vec g}/\partial M_o)\times\efbold\;\right)\;$
  have been
  omitted in (\ref{9} - \ref{10}), because these terms vanish
  identically. (For technical details see the Appendix to Efroimsky (2005a).)

\subsubsection{\textbf{The Approximation}}

 To obtain the first-order (over $\,\mubold\,$) secular parts of the osculating elements,
 we shall carry out two operations:\\

 \textbf{1.} ~~First, we shall throw out $\,O(\mubold^2)\,$ contribution to $\,``\Delta
 {\cal H}"\,$ and shall assume that preservation of the first-order terms and neglect of
 the second-order ones in the equations makes these equations render solutions valid in
 the first order. This assumption should remain valid for some interval of time, an
 interval whose actual duration can be determined only through accurate numerical
 simulation. In our analytical developments we shall hope that this interval is
 sufficiently long.\footnote{~In (\ref{9} - \ref{14}), the quadratic in $\,\mubold\,$
 terms in the right-hand side will be of order $\,|\mubold|^2/n\,$. According to Ward
 (1973), the range of values of $\;|\mubold|\;$ for Mars hardly ever exceeded $\;10^{-3}
 \,yr^{-1}$
% , and is typically less than $\;10^{-4}\;$
. The value of $\;n\;$, for the Martian satellites, is of order one
 day$^{-1}$. If we now look, for example, at (\ref{12}), we shall see that the quadratic
 in $\,\mubold\,$ terms surely cannot contribute to $\;d\inc/dt\;$ more than an angular
 degree over a million of years, and are quite likely to remain insignificant over dozens
 of millions years. Whether these terms may be neglected at timescales longer than $100$
 millions of years -- should be checked through a comparison of our semianalytical
 model with a comprehensive numerical simulation. As demonstrated in Gurfil, Lainey, \&
 Efroimsky (2007), the answer to this question turns out to be positive: the model remains
 surprisingly exact for, at least, $\,20$ Myr.}\\

 \textbf{2.} ~~Second, we shall substitute both the disturbing function
 $\;\left(\,-\,``\Delta {\cal H}"\,\right)\;$ and the other precession-generated (i.e.,
 $\;\mubold$-dependent) terms with their orbital averages. To be more exact, the
 precession rate $\,\mubold\,$ and each of the elements will be considered as a function
 of the true anomaly $\,\nu\,$, and will be expanded into a Fourier integral which will
 then be split into two pieces -- an integral over the band of frequencies less than the
 orbital frequency and an integral over the higher frequencies:
 \ba
 C_j\;=\;\bra\,C_j\,\ket\;+\;C_j^\heartsuit\;\;\;,\;\;\;\;\;
 \mubold\;=\;\bra\,\mubold\,\ket\;+\;\mubold^{\,\heartsuit}
 \label{15}
 \ea
 The first piece ($\,\bra \mubold \ket\,$ or $\,\bra C_j\ket\,$) will be regarded as the
 secular part, while the second piece ($\,\mubold^\heartsuit\,$ or $\,C_j^\heartsuit\,$)
 will be averaged out. One of the outcomes of this treatment will be that the left-hand
 side of the averaged planetary equations will now contain not the time derivatives of the
 elements but the derivatives of their secular parts. To understand the structure of the
 averaged right-hand sides, let us consider some product $\;A(\nu)\,B(\nu)\;$, where $\,A
 \,$ and $\,B\,$ denote some of the orbital elements or the projection $\,\mu_{\perp}\,$
 of the vector $\,\mubold\,$ onto the instantaneous normal to the satellite orbit (this
 projection will appear many times in our analysis of the planetary equations):
 \ba
 A\;B\;=\;\left(\bra\,A\,\ket\;+\;
 A^\heartsuit\right)\,\left(\bra\,B\,\ket\;+\;
 B^\heartsuit
 \right)\;=\;\bra A\;B \ket\;+\;
 \left(A\;B \right)^\heartsuit\;\;\;.
 \label{16}
 \ea
 The secular and high-frequency components of this product will read, correspondingly, as
 \ba
 \bra A\;B \ket\;\equiv\;
 \bra\,A\,\ket\;\;\bra\,B\,\ket\;+\;
 \bra A^\heartsuit\,B^\heartsuit \ket
 ~~~~~~~~~~~~~~~~~~~~~~~~~~~~~~~
 \label{17}
 \ea
 and
 \ba
 \left(A\;B \right)^\heartsuit\;\equiv\;
 \bra\,A\,\ket\;\,B^\heartsuit\;+\;
 \bra\,B\,\ket\;A^\heartsuit\;+\;\left(
 A^\heartsuit\,B^\heartsuit \;
 -\;\bra A^\heartsuit\,B^\heartsuit \ket
 \right)\;\;\;.
 \label{18}
 \ea
 An obvious circumstance is that the secular part of the product consists not only of the
 product of the secular parts of the multipliers but also of the term $\;\bra A^\heartsuit
 \,B^\heartsuit \ket\;$ containing resonances. A less evident but crucially important
 circumstance is that, technically, the above separation of timescales is never
 implemented exactly (unless one deals from the very beginning with the Fourier expansions
 of all the functions involved). Therefore, the (imperfectly calculated) high-frequency
 parts $\;A^\heartsuit\;$, $\;B^\heartsuit\;$, and $\; \left(A\;B \right)^\heartsuit\;$
 are unavoidably contaminated with the lower-frequency modes, modes whose effect may
 considerably accumulate at large times and exert ``back-reaction" upon the secular part
 of the product. (Laskar 1990)

\subsubsection{\textbf{The planetary equations
for the first-order secular parts of the osculating elements}}

Naively, the afore proposed approximation will lead us to a new
system of planetary equations. It will be identical to the system
(\ref{9} - \ref{14}), except that now the letters
$\;a\,,\;e\,,\;\omega\,,\;\Omega\,,\;\inc\,,\;M_o\;$ will denote
not the osculating elements but their secular parts. Similarly,
$\;\mubold\;$ will now stand for the secular part of the
precession rate. The Hamiltonian will now be substituted with
 \begin{eqnarray}
 \nonumber
 \Delta {\cal H}^{(eff)}\;=\;-\;\left[\;\bra R_{oblate} \ket\;+\;
 \bra\, \mubold\cdot(\efbold\times {\bf {\vec{g}}})\,\ket \;\right]\;=
 ~~~~~~~~~~~~~~~~~~~~~~~~~~~~\\
 \label{19}\\
 \nonumber
 -\;\frac{G\,m\,J_2}{4}\;\;\frac{\rho_e^2}{a^3}\;\;
 \frac{3\,\cos^2i\,-\,1}{\left(1\,-\,e^2\right)^{3/2}}\,-\,
 \sqrt{G\,m\,a\,\left(1\,-\,e^2\right)}\;\;\left(\,
 \mu_1  \;\sin i\;\sin \Omega\,-\, \mu_2  \;\sin i\;\cos \Omega
 \,+\, \mu_3 \;\cos i \,\right)\,~~
 \end{eqnarray}
where, once again, all letters denote not the appropriate
variables but their averages.

By doing so, we would, however, ignore that the
$\;\mubold$-dependent terms in (\ref{9} - \ref{14}) contain
products of high-frequency quantities (such as, for example, the
product of the true-anomaly-dependent expression $\;\;(\partial
\efbold/\partial \omega)\times {\bf\vec g}\,-\, (\partial {\bf
\vec g}/\partial \omega)\times\efbold\;$ by the high-frequency
part of $\;\mubold\;$ in formula (\ref{10}) ). Averages of such
products will contribute to the secular parts of the right-hand
sides of the approximate planetary equations, as in the example
(\ref{17}). (As we shall see below, these inputs will be due to
the commensurabilities between the orbital motion of the satellite
and the short-term nutations of the primary.) Keeping this in
mind, we should approximate the exact planetary equations rather
with the following system:
 \ba
 \frac{da}{dt}\,=\,
% 0\;\;\;,\;\;\;\;\;\;\;\;\;\;\;\;\;~~~~~~~~~~~~~~~~~~~~~~~~~~~~~~~~~~~~
 \frac{2}{n a}\,\left[
%\frac{\partial\left(\,-\,\Delta {\cal H}^{(eff)}\right)
% }{\partial M_o}
 \;-\;\langle \;\dotmubold\; \left(\efbold\,\times\,
 \frac{\partial \efbold}{\partial M_o}  \right)\; \rangle\;\right]
 \;\;\;,\;~~~~~~~~~~~~~~~~~~~~~~~~~~~~~~~~~~~~~~~~~~~~~~~~~~~~~~~~
 \label{20}
  \ea
  ~\\
 ~\\
  \begin{eqnarray}
  \nonumber
 \frac{de}{dt}\,=\,
 \frac{1-e^2}{n\,a^2\,e}\;\;\left[\;
 \;-\;\langle \;\dotmubold\; \left(\efbold\,\times\,
 \frac{\partial \efbold}{\partial M_o}  \right)\; \rangle \;\right]~~
 \,~~~~~~~~~~~~~~~~~~~~~~~~~~~~~~~~~~~~~~~~~~~~~~~~~~~
 \\
 \nonumber\\
 \label{21}
 \\
 \nonumber\\
 \nonumber
 -\;\frac{(1\,-\,e^2)^{1/2}}{n\,a^2\,e} \;\left[\;
% \frac{\partial \left(\,-\,\Delta {\cal H}^{(eff)} \right) }{\partial \omega}\,~ +\,
 \langle~ \mubold\cdot
 \left(\frac{\partial{\efbold}}{\partial \omega}\times
 {\bf{\vec g}}\,-\,{\efbold}\times
 \frac{\partial{\vec{\bf g}}}{\partial \omega}\right)~\rangle
 \;-\;\langle \; \dotmubold\; \left(\efbold\,\times\,
 \frac{\partial \efbold}{\partial \omega}  \right)\;
 \rangle\;\right]~~,~~~~~~~~
  \end{eqnarray}
 ~\\
 ~\\
  \begin{eqnarray}
 \nonumber
 \frac{d\omega}{dt}\,=~
 \frac{\;-\,\cos \inc
}{n a^2(1-e^2)^{1/2}\, \sin \inc }\left[ \frac{\partial
\left(\,-\,\Delta {\cal H}^{(eff)} \right) }{\partial \inc } \,~ +
~ \langle~\mubold\cdot
 \left(\frac{\partial{\efbold}}{\partial \inc}\times
 {\bf{\vec g}}\,-\,{\efbold}\times
 \frac{\partial{\vec{\bf g}}}{\partial \inc}\right)\,~\rangle
 \;-\;\langle \; \dotmubold\; \left(\efbold\,\times\,
 \frac{\partial \efbold}{\partial \inc}  \right)\;
 \rangle\;
  \right]~~~~
  \ea
  \ba
 \nonumber\\
 \nonumber\\
 \label{22}\\
 \nonumber
 +\,~\frac{(1-e^2)^{1/2}}{n\,a^2\,e}\;
 \left[\;\frac{\partial \left(\,-\,\Delta {\cal H}^{(eff)} \right) }{\partial e}\,
+ ~ \langle~\mubold\cdot
 \left(\frac{\partial{\efbold}}{\partial e}\times
 {\bf{\vec g}}\,-\,{\efbold}\times
 \frac{\partial{\vec{\bf g}}}{\partial e}\right)~\rangle
 \;-\;\langle \; \dotmubold\; \left(\efbold\,\times\,
 \frac{\partial \efbold}{\partial e}  \right)\;
 \rangle\;
 \;\right]\;\;,~~~~~~
  \end{eqnarray}
 ~\\
 ~\\
 ~\\
 \ba
 \nonumber
 \frac{d \inc
 }{dt}\;=~ \frac{\cos \inc}{n a^2\,(1 - e^2)^{1/2} \sin
 \inc}
 \left[
 \; \langle~\mubold\cdot
 \left(\frac{\partial{\efbold}}{\partial \omega}\times
 {\bf{\vec g}}\,-\,{\efbold}\times
 \frac{\partial{\vec{\bf g}}}{\partial \omega}\right)~\rangle
 \;-\;\langle \; \dotmubold\; \left(\efbold\,\times\,
 \frac{\partial \efbold}{\partial \omega}  \right)\;
 \rangle\;
 \right]\,-\;\;\;
 ~~~~~~~~~~~~~~~\\
 \nonumber\\
 \label{23}\\
 \nonumber\\
 \nonumber
 \frac{1}{na^2\,(1-e^2)^{1/2}\,\sin \inc
 }\,\left[\,\frac{\partial \left(\,-\,\Delta {\cal H}^{(eff)} \right) }{\partial
 \Omega}\;+ \; \langle~\mubold\cdot
 \left(\frac{\partial{\efbold}}{\partial \Omega}\times
 {\bf{\vec g}}\,-\,{\efbold}\times
 \frac{\partial{\vec{\bf g}}}{\partial \Omega}\right)~\rangle
 \;-\;\langle \; \dotmubold\; \left(\efbold\,\times\,
 \frac{\partial \efbold}{\partial \Omega}  \right)\;
 \rangle\;
 \,\right]\;{,}\;\;\;
  \ea
 ~\\
 ~\\
  \ba
\frac{d\Omega}{dt} = \frac{1}{n a^2\,(1-e^2)^{1/2}\,\sin \inc
}\,\left[\,\frac{\partial \left(\,-\,\Delta {\cal H}^{(eff)}
\,\right) }{\partial \inc }\, + \,\langle \,\mubold\cdot
 \left(\frac{\partial{\efbold}}{\partial \inc}\times
 {\bf{\vec g}}\,-\,{\efbold}\times
 \frac{\partial{\vec{\bf g}}}{\partial \inc}\right)\,\rangle
 \,-\,\langle \, \dotmubold\, \left(\efbold\,\times\,
 \frac{\partial \efbold}{\partial \inc}  \right)\,
 \rangle\,
 \right]\,,\;\;
 \label{24}
  \ea
 ~\\
 ~\\
  \begin{eqnarray}
  \nonumber
\frac{dM_o}{dt}\,=\,\;-\,\;\frac{1\,-\,e^2}{n\,a^2\,e}\,\;
\left[\;\frac{\partial \left(\,-\,\Delta {\cal H}^{(eff)} \right)
 }{\partial e }~ + ~
 \langle~\mubold\cdot
 \left(\frac{\partial{\efbold}}{\partial e}\times
 {\bf{\vec g}}\,-\,{\efbold}\times
 \frac{\partial{\vec{\bf g}}}{\partial e}\right)~\rangle
 \;-\;\langle \; \dotmubold\; \left(\efbold\,\times\,
 \frac{\partial \efbold}{\partial e}  \right)\;
 \rangle
 \;\right]\\
 \nonumber\\
 \label{25}\\
 \nonumber\\
 \nonumber
 -\;~\frac{2}{n\,a}\,\left[\;\frac{\partial \left(\,-\,\Delta {\cal H}^{(eff)} \right)
 }{\partial a }~ + ~ \langle ~\mubold\cdot
 \left(\frac{\partial{\efbold}}{\partial a}\times
 {\bf{\vec g}}\,-\,{\efbold}\times
 \frac{\partial{\vec{\bf g}}}{\partial a}\right)~\rangle
 \;-\;\langle \; \dotmubold\; \left(\efbold\,\times\,
 \frac{\partial \efbold}{\partial a}  \right)\;
 \rangle\;
 \right]
 \;\;,~~~~~
  \end{eqnarray}
  ~\\
the angular brackets denoting the secular parts. In (\ref{16}),
(\ref{17}) and (\ref{19}) we took into account the fact that the
averaged and truncated Hamiltonian (\ref{15}) depends neither on
$\;M_o\;$ nor on $\;\omega\;$.

It should be emphasised that in this subsection and hereafter
\emph{the symbols} $\;a\,,\;e\,,\;\omega\,,\;\Omega\,,\;\inc\,,$
$M_o\,,\;\mubold\;$ \emph{stand not for the exact values but for
the mean values of the appropriate variables}. A mean value of an
element is understood to include the secular and long-period
parts, the short-period components being averaged out.

The case of uniform planetary precession ($\;\mubold\,=\,const\;$)
was studied in Efroimsky (2005a). In that case, the terms
containing $\;\dotmubold\;$ evidently vanish. Besides, it turns
out that, {\underline{for constant $\;\mubold\;$}}, the mean
values of the other $\;\mubold$-dependent terms, except one,
vanish too:
  ~\\
 \ba
 \mubold\;\cdot\;\langle\;\left(\; \frac{\partial \efbold}{\partial C_j}\;\times\;{\bf \vec
g}\;-\;\efbold\;\times\;\frac{\partial {\bf \vec g}}{\partial
C_j}\;\right)\;\rangle\;=\;0\;\;\;~,~~~~~~~~~
C_j\;=\;e\,,\;\Omega\,,\;\omega\,,\;\inc\,,\;M_o\;\;\;,
~~~~~~~~~~~~~~~~~~~~
 \label{26}
 \ea
~\\
  \ba
 \mubold\,\cdot\,\langle\;\left(\;\frac{\partial \efbold}{\partial a}\,\times\,{\bf \vec
g}\,-\,\efbold\,\times\,\frac{\partial {\bf \vec g}}{\partial
a}\;\right)\;\rangle\,=\,\mubold\,\cdot\,\left(\;\frac{\partial
\efbold}{\partial a}\,\times\,{\bf \vec
g}\,-\,\efbold\,\times\,\frac{\partial {\bf \vec g}}{\partial
a}\;\right) \,=\,\frac{3}{2}\;\;
\mu_{\perp}\;\;\sqrt{\frac{G\;m\;\left(1\;-\;e^2\right)}{a}}\;\;\;,~~
 \label{27}
 \ea
 where
 \ba
 \nonumber
 \mu_{\perp}\;\equiv\;\mu_1\;\sin \inc\;\sin \Omega\;-
 \;\mu_2\;\sin \inc\;\cos \Omega\;+\;\mu_3\;\cos \inc~~~~~~~~~~~~~~~~\\
 \label{}\\
 \nonumber
 =\;{\dot{I}}_p\;\sin \inc\;\sin \Omega\;-
 \;{\dot{h}}_p\;\sin I_p\;\sin \inc\;\cos \Omega\;+
 \;{\dot{h}}_p\;\cos I_p\;\cos \inc
 \label{28}
 \ea
is the projection of the planets' precession rate $\;{\mubold}\;$
onto the instantaneous normal to the satellite's orbit.
\footnote{~Here $\,\mu_{\perp}\,$ is expressed in the basis
$\;{\bf\hat x}\,,\;{\bf\hat y}\,,\;{\bf\hat z}\;$ associated with
the planet's equator of date.
% $\,\mu_1\,\equiv\,\mubold\,\cdot\,{\bf{\hat x}}\,,\;\;\;
%   \mu_2\,\equiv\,\mubold\,\cdot\,{\bf{\hat y}}\,,\;\;\;
%   \mu_3\,\equiv\,\mubold\,\cdot\,{\bf{\hat z}}\,$.
   Unit vector $\;{\bf{\hat z}}\;$ is
perpendicular to the equator of date, while $\;{\bf{\hat x}}\;$ is
pointing along the line of the ascending node of the equator of
date on the equator of epoch; therefore, the components
$\;\mu_{j}\;$ are given by (\ref{3}).
 % through the longitude of the node, $\;h_p\;$, and the inclination,
 % $\;I_p\;$, of the equator of date relative to that of epoch. If not
 % for the precession of the planetary orbit about the Sun, calculation
 % of $\;h_p\,,\;I_p\;$ and their time derivatives would be trivial. For
 % example, in the case of an symmetrical oblate planet the inclination
 % would stay unchanged, while the node would be moving at a constant rate.
 % In reality, computation of $\;h_p\,,\;I_p\;$ is laborious, and is
 % explained below in subsection 2.2.2 and, in more detail, in the Appendix.
 }

Hence, in this approximation and under the assumption of constant
$\,\mubold\,$, in order to compute the secular parts of the
orbital elements, it is sufficient to amend the Hamiltonian with
the $\;\mubold$-dependent addition and to ignore all the other
$\;\mubold$-dependent terms except the one given by (\ref{27}).
This will no longer be the case for variable precession, i.e., for
time-dependent $\;\mubold\,$. Section 2 of our article will
address itself to calculation of the secular parts (\ref{26} -
\ref{28}) in the case of time-dependent $\;\mubold\;$.

 \section{Equations for the first-order secular parts\\
          of the osculating elements.}

\subsection{Two Fourier expansions of the precession spectrum}

Precession of the planetary spin axis has a continuous spectrum
that spans from the polar wander and the fastest nutations to the
Chandler wobble to the long-term variations whose time scales go
all way to billions of years. When the planet has a sufficiently
massive moon capable of influencing the planetary precession, the
rate of this precession, $\,\mubold\,$, should be regarded as a
function not only of time but also of the position of the
satellite. We shall be interested, however, in the situation where
the satellites are small and do not considerably influence
rotation of their primary (while rotation variations of the
primary still may affect the satellite orbits). This is, for
example, the case of Mars whose tiny satellites affect its
precession only in a very high order (Laskar 2004). Under these
circumstances, it is fair to treat the precession rate as a
function of time solely:
 \ba
 \mubold(t)\;=\;\int_{0}^{\infty}\;\left[
 \;\mubold^{(s)}(u)\;\sin({ut})\;+\;
 \mubold^{(c)}(u)\;\cos({ut})\;\right]\;du\;\;\;,
 \label{29}
 \ea
$\;u\;$ standing for the angular frequency. In what follows, it
will be convenient to describe the evolution not in terms of time
but via the true anomaly $\,\nu\,$ of the satellite. For our
present purposes, it will be advantageous to express the
precession rate as function of the satellite's true anomaly:
 \ba
 \mubold(\nu)\;=\;\int_{0}^{\infty}\;\left[
 \;\mubold^{(s)}(W)\;\sin({W\nu})\;+\;
 \mubold^{(c)}(W)\;\cos({W\nu})\;\right]\;dW\;\;\;.
 \label{30}
 \ea
$W\,$ being the circular ``frequency" related to the true
 anomaly $\;\nu\;$. Evidently,
 $\;\mubold(t)\,\;\mubold(\nu)\,,\;\mubold(u)\,,\;$ and
 $\;\mubold(W)\;$ are four different functions. We nevertheless
 denote them with the same notation $\;\mubold(\,.\,.\,.\,)\;$
because the argument will always single out which particular
function we mean. The interconnection between functions $\;\mubold
(\nu)\;$ and $\;\mubold(t)\;$ is given by
 \ba
 \nonumber
 \mubold(t)\;=\;\mubold(\nu)\,|\,_{_{\nu\,=\,\int n\,dt}}\;\;\;.
 \label{}
 \ea
The interconnection between the Fourier components is less
obvious. However, it simplifies under the assumption of vanishing
eccentricity and slowly-changing semimajor axis:
 \ba
 \mubold(W)\;\approx\;n\,\mubold(u)\,{\mid}_{\textstyle \left.\,\right._{\small
 W\,=\,u/n}}\;\;\;,\;\;\;\;\;\;\;n\;\equiv\;(G\,m)^{1/2}\;a^{-3/2}\;\;\;.
 \label{491}
 \label{31}
 \ea
 A rigorous relation to be used below is\footnote{~This relation immediately
 follows from the well known equality $\;dM\;\left( 1\,+\,e\, \cos \nu\right)^2\;
 =\;d\nu\; \left( 1\,-\,e^2 \right)^{3/2} \;$, one upon which also the
averaging rule
 (\ref{rule}) is based.}:
  \ba
 \frac{d\mubold
 (\nu)}{d\nu}\;=\;\frac{d\mubold (t)}{dt}
 \left(\frac{\partial t}{\partial\nu}\right)_{{a,\,e,\,\inc,\,\omega,\,\Omega,\,
 M_o}} =\;\,\dotmubold\,\left(\frac{\partial t}{\partial M}\right)_{{a,\,.\,.\,.}}
 \left(\frac{\partial M}{\partial\nu}\right)_{{a,\,.\,.\,.}}=\;
 \frac{\dotmubold\left(1 - e^2\right)^{3/2}}{n\left(1 + e\,\cos\nu
 \right)^2}
 \;\;.\;\;\;
 \label{491}
 \label{32}
 \ea
 ~\\

 \subsection{The role of the $\;\langle\,\mubold\cdot\left(\,.\,.\,.\,\right)\,\rangle\;$
 and $\;\langle\,\dotmubold\cdot\left(\,.\,.\,.\,\right)\,\rangle\;$ terms}

These terms, ignored in the literature hitherto, implement the
subtle influence of the planet's orbit precession upon its
satellites' motion. The physical content of this effect is as
follows: first, the precession of the planetary orbit slowly alters
the solar torque acting on the planet; second, the variations of
this torque entail changes in the planetary spin axis' precession;
and, finally, third: these changes influence the satellites' orbits.
This three-step interaction is extremely weak; still, its effect may
accumulate over very long periods of time.

\subsubsection{The $\;\langle\,\mubold\cdot\left(\,.\,.\,.\,\right)\,\rangle$ terms}

To illustrate the role of commensurabilities between the satellite
orbital motion and the planetary nutations, let us consider the
average $\;\bra\, \;\mubold\cdot\left(\;(\partial \efbold/\partial
e)\times {\bf\vec g}\,-\, (\partial {\bf \vec g}/\partial
e)\times\efbold\;\right) \;\,\ket\;$ emerging in the equation
(\ref{22}) for $\;d\omega/dt\;$ and in the equation (\ref{25}) for
$\;dM_o/dt\;$: As shown in section A.4 of the Appendix to
Efroimsky (2004),
 \ba
 \mubold\cdot\left(\frac{\partial \efbold}{\partial e}\;\times\;{\bf \vec
 g}\;-\;\efbold\;\times\;\frac{\partial {\bf \vec g}}{\partial
 e}\right)\;=\;
  -\;\,\mu_{\perp}\;\,\frac{n\,a^2\;\left(3\,e\,+\,2\,\cos \nu\,+\,e^2\,\cos \nu
  \right)}{\left(1\,+\,e\,\cos\nu\right)\;\sqrt{1\;-\;e^2}}~~~,~~~~~~~~~~~~
 \label{33}
 \ea
 $\mu_{\perp}\,$ being given by (\ref{28}). By virtue of (\ref{rule}) and
 (\ref{30}), its secular part at some $\,\nu\,$ will be:
 \ba
 \nonumber
 \bra\,\mubold\cdot\left(\frac{\partial \efbold}{\partial e}\times{\bf \vec
 g}-\efbold\times\frac{\partial {\bf \vec g}}{\partial
 e}\right)\ket\,=\,-\,\frac{1-e^2}{2\pi}\,na^2
 \int_{\nu'=-\pi}^{\nu'=\pi}\mu_{\perp }(\nu+\nu\,')\,
  \frac{3e+2\,\cos (\nu+\nu\,')+ e^2\,\cos (\nu+\nu\,') }{\left(1\,+
  \,e\;\cos (\nu+\nu\,')
  \right)^3}d\nu\,'=
  \ea
  ~\\
  \ba
 \nonumber
 -\,\frac{1-e^2}{2\,\pi}\,n\,a^2\,\int_{0}^{\infty}d
 W \int_{-\pi}^{\pi}d\nu\,'\left[\,
 \mu^{(s)}_{\perp}(W)\;\sin({W(\nu +\nu\,')})\,+\,
 \mu^{(c)}_{\perp}(W)\;\cos(W(\nu +\nu\,'))
 \,\right]\,\left[\,\left(2+e^2\right)\,\cos(\nu \right.
 \ea
 ~\\
 \ba
 \nonumber
 \left.+\,\nu\,')\;+\;\left(\;-\;3\;e\;-\;\frac{5}{2}\;e^3\right)\;\cos 2(\nu+\nu\,')\;
 +\;3\;e^2\;\cos 3(\nu+\nu')\;-\;\frac{5}{2}\;e^3
 \;\cos 4(\nu+\nu')\;+\;O(e^4) \;\right]\;=
 \ea
 \\
 % ~~~.~~~~~~~~~~~~~~~~~~~~~
 %  \label{34}
 % \ea
 % After evaluation of the integrals over $\;\nu\,'\;$, this will read:
 % \ba
 % \bra\,\mubold\cdot\left(\frac{\partial \efbold}{\partial e}\;\times\;{\bf \vec
 % g}\;-\;\efbold\;\times\;\frac{\partial {\bf \vec g}}{\partial
 % e}\right)\,\ket\,
 \ba
 -\,\frac{1-e^2}{2\,\pi}\,na^2
 \left[
 \left(2+e^2\right)\,\mu_{\perp}^{(c)}(1)\,+\,
 \left(-3e-\,\frac{5}{2}\,e^3\right)\,\mu_{\perp}^{(c)}(2)\,+\,
 3e^2\,\mu_{\perp}^{(c)}(3)\,-\,
 \frac{5}{2}\,e^3\,\mu_{\perp}^{(c)}(4)\,+\,O(e^4)
 \right]~~~.~~~~
 \label{34}
 \ea
 $W\;$ being the angular ``frequency" related to the true anomaly $\;\nu\;$, as in
 equation (\ref{30}). Not surprisingly, the integral over $\,W\,$ has been reduced
 to an infinite sum over the discrete values $\,W\,=\,1,\,2,\,3,\,4,\,.\;.\;.\,$
 corresponding to commensurabilities between the orbital frequency of the satellite
 and the nutational frequencies of the oblate planet.\footnote{~It should be
 emphasised that (\ref{34}) was obtained by a certain approximation: the averaging
 ignored the back-reaction of the short-period motions upon the long-period ones
 (i.e., it ignored the fact that, after each orbital period, the satellite does
 not return to exactly the same point it started); for example, it was assumed
 that the elements $\,e\,$ and $\,a\,$ remained constant during the integration
 over $\;\nu\,'\;$ from $\,0\,$ through $\,2\pi \,$.} The main resonant input
 comes from the principal commensurability $\;W\,=\,1\;$, i.e., from the nutation
 mode resonant with the orbit. The higher-order resonant inputs originate from the
 faster nutations characterised by $\;W\,=\,2\,,\;3,\,\;4\;.\;.\;.\;\;$
 In equations (\ref{20} - \ref{25}), almost all terms proportional to $\,\mubold\,$
 produce such resonances. At the time when we are writing this paper, our knowledge
 about the fast nutations and polar wonder of Mars is yet very limited, and we shall
 not venture to offer quantitative estimates of the time scale over which the effect
 of these resonances upon the satellite orbit becomes considerable.

 Slower than $\,W\,=\,1\,$ variations of $\,\mubold\,$ bring no nonresonant contributions
 into the average of the right-hand side of (\ref{33}). It can be shown that none
 of the $\;\langle\,\mubold\cdot\left(\,.\,.\,.\,\right)\,\rangle\;$ terms emerging in
 (\ref{20} - \ref{24}) yield a nonresonant input. (Hence, (\ref{26}).) For these reasons,
 in the rest of this paper, the terms $\;\langle\,\mubold\,\left(\,.\,.\,.\,\right)\,
 \rangle\;$ will be omitted.

 \subsubsection{The $\;\langle\,\dotmubold\cdot\left(\,.\,.\,.\,\right)\,\rangle$ terms}

 Let us consider, as an example, the term
 $\,\langle\,\dotmubold\cdot\left(\,-\,\efbold\,\times({\partial {\efbold}}/{
 \partial \omega})\,\right)\,\rangle\,$ showing up in the right-hand sides of
 equations (\ref{21}) and (\ref{23}). We have from Appendix A11 of Efroimsky (2004):
 \ba
%
% THIS EXPRESSION HAS UNDERGONE
%
% A COMPREHENSIVE CHECK
%
 \dotmubold\;\cdot\;\left(\;-\;\efbold\;\times\;\frac{\partial \efbold}{\partial
 \omega}\;\right)\;=\;-\;\dot{\mu}_{\perp}\;\;a^2\;\;\frac{\left(1\;-\;e^2\right)^2}{\left(1\;+\;e\;\cos
\nu\right)^2}\;\;\;\;.~~~~~~~~
 \label{35}
 \ea
 Just as in the preceding example (\ref{34}), orbital averaging of this expression would
 yield resonant terms entailed by commensurabilities between the orbital frequency of the
 satellite and the fast variations of $\,\dotmubold\,$. For the reasons explained above,
 here we omit these contributions. However, in distinction from the $\;\langle\,\mubold
 \cdot\left(\,.\,.\,.\,\right)\,\rangle\;$ terms, some of the $\;\langle\,\dotmubold\cdot
 \left(\,.\,.\,.\,\right)\,\rangle\;$ do have nonresonant components. For example, the mean
 part of (\ref{35}) will be finite even for a constant $\,\dotmubold
 \,$:
 \ba
 \bra\,\dotmubold\cdot
 \left(-\efbold\times\frac{\partial \efbold}{\partial
 \omega}\right)\,\ket\,=\,
-\,\dot{\mu}_{\perp}a^2\left(1-e^2\right)^2\,\frac{\left(1-e^2\right)^{3/2}}{2\;\pi}\,
\int_{-\pi}^{\pi} \,\frac{d\nu
 }{\left(1+e\,\cos
\nu\right)^4}\,=\,
 -\,\dot{\mu}_{\perp}\,\frac{a^2}{2}\,\left(2+3\,e^2
 \right)~~,~~
 \label{36}
 \ea
 the superscript dot denoting a time derivative taken in the frame co-precessing with the
 equator of date. In other words, $\;\dot{\mu}_{\perp}\;$ is, by definition, not a full
 time derivative but its projection onto the instantaneous normal to the satellite's
 orbit. So defined $\;\dot{\mu}_{\perp}\;$ contains only derivatives of $\;\mu_j\;$ but
 not of the angles:
 \ba
 \dot{\mu}_{\perp}\;=\;\dot{\mu}_1\;\sin\inc\;\sin\Omega\,-\,\dot{\mu}_2\;\sin\inc\;\cos
 \Omega\,+\,\dot{\mu}_3\;\cos\inc\;\;\;.~~~~~~~~~~~~~~~~~~~~~~~~~~~~~~~~~~~~~~~~~~~~~~~~
 ~~~~~~~~~~~
 \label{37}
 \ea
 As shown in the Appendix below, $\;\dot{\mu}_{\perp}\;$ can be expressed via the
 longitude of the node, $\,h_p\,$, and the inclination, $\,I_p\,$, of the equator of
 date relative to the one of epoch:
 \ba
 \nonumber
 {\dot{\mu}}_{\perp}\,=\,
 {\ddot{I}}_p\,\sin \inc\;\sin \Omega\,-
 \,\left(\,
 {\ddot{h}}_p\,\sin I_p\,+\,{\dot{h}}_p\,{\dot{I}}_p\,\cos I_p
 \,\right)\,\sin \inc\,\cos \Omega\,+
 \,\left(\,
 {\ddot{h}}_p\,\cos I_p\,-\,{\dot{h}}_p\,{\dot{I}}_p\,\sin I_p
 \,\right)\,\cos \inc\;\;\;\;\;\;\;\;\;\;\;\;\;\;
 \label{}
 \ea
 \ba
 {\;}^{\;\;}~~~\approx\;{\ddot{h}}_p\,\left(\,
 -\,\sin I_p\;\sin\inc\;\cos\Omega\;+\;\cos I_p\;\cos\inc
 \,\right)~~~.~~~~~~~~~~~~~~~~~~~~~~~~~~~~~~~~~~~~~~~~~~~~~~~~~~~~~~~~~~~~
 \label{38}
 \ea
 The quantities $\;h_p\,,\;I_p\;$ and their time derivatives can be calculated from
 integration of the Colombo equation of spin precession in inertial space,
 \ba
 \frac{d{\bf\hat k}}{dt}\;=\;\alpha\;\left({\bf{\hat{k}}}\cdot{\bf{\hat{n}}}
 \right)\,\left({\bf{\hat{k}}}\times{\bf{\hat{n}}}  \right)\;\;\;,
 \label{39}
 \ea
${\bf\hat k}\;=\; \left(\;\sin I_p\,\sin h_p\;\;,\;\;\;-\,\sin
I_p\,\cos h_p
  \;\;,\;\,\;\cos I_p\;\right)^{^T}
\;$ being a unit vector pointing along the major-inertia axis of
the planet, and $\;{\bf\hat n}\;= \left(\;\sin I_{orb}\,\sin
\Omega_{orb}\;\;,\;\;\;-\,\sin I_{orb}\,\cos \Omega_{orb}
  \;\;,\;\,\;\cos I_{orb}\;\right)^{^T}
\;$ being a unit normal to the planetary orbit plane defined
(relative to some fiducial plane) through the inclination
$\;I_{orb}\;$ and longitude of the node $\;\Omega_{orb}\;$. The
constant (or, better to say, the slowly-varying factor)
$\;\alpha\;$ is given by
 \ba
 \alpha\;\equiv\;\frac{3\;n^2_p}{2\;s\;\left(1\,-\,e_p^2\right)^{3/2}}\;\,
 \frac{C\;-\;\frac{\textstyle A\,+\,B}{\textstyle 2} }{C}\;\;
 \label{alpha}
 \label{40}
 \ea
where $\;n_p\;$, $\;e_p\;$, $\;s\;$, and $\;A\,,\;B\,,\;C\;$ are the
mean motion, the eccentricity, the spin angular velocity, and the
moments of inertia of the planet. (As ever,
$\;C\,\geq\,B\,\geq\,A\;$.) Even in the relatively simple case of
$\;C\,>\,B\,=\,A\;$, the planet's axis of rotation does not describe
a circular cone because the unit normal to the planet's orbit,
$\;\bf\vec n\;$, is subject to variations caused by the precession
of the planet's orbit about the Sun. While integration of the
Colombo equation is explained below in the Appendix, here we would
emphasise that this equation describes the evolution of planetary
spin only under a very strong assumption of this spin being
principal, i.e., in neglect of the Chandler wobble and polar wander.

% \pagebreak

\section{Evolution of the elements in the leading order of $\,e\,$}
%~\\

\subsection{\textbf{The semimajor axis and the eccentricity}}

 As explained in subsection 2.2.1, the $\;\langle\,\mubold\cdot(\,.\,.\,.\,)\,\rangle$
 terms may be omitted. The expressions for the orbital averages
 of the $\,\dotmubold$-dependent terms, derived in the Appendix, have the form:
 \ba
 \bra\;\dotmubold\;\cdot\;\left(\;-\;\efbold\;\times\;
 \frac{\partial {\efbold}}{\partial
 M_o}\;\right)\;\ket\;=\;
 -\;\dot{\mu}_{\perp}\;a^2\;\sqrt{1\;-\;e^2}\;\;\;,~~~~~~~~~~~~\;\;\;\;\;
 \label{41}
 \ea
 \ba
 \bra\;\dotmubold\;\cdot\;\left(\;-\;\efbold\;\times\;\frac{\partial \efbold}{\partial
 \omega}\;\right)\;\ket\;=\;
  -\;\dot{\mu}_{\perp}\;\;a^2\;\left(\,1\;+\;\frac{3}{2}\;e^2\right)
  \;\;.~~~~~~~~~~~~~~~~
 \label{42}
 \ea
 Here
 \ba
 \mu_{\perp}\;\equiv\;\mubold\,\cdot\,\wbold\;=\;\mu_1\;\sin \inc\;\sin \Omega\;-
 \;\mu_2\;\sin \inc\;\cos \Omega\;+\;\mu_3\;\cos \inc\;\;\;,
 \label{43}
 \ea
 where the unit vector
 \ba
 \nonumber
 {\wbold}\;=\;{\bf\hat{x}}\;\sin \inc\;\sin \Omega\;-
 \;{\bf\hat{y}}\;\sin
 \inc\;\cos \Omega\;+\;{\bf\hat{z}}\;\cos
 \inc\;\;~~~~~~~~~~~~~~~~~~~
 \label{}
 \ea
 is the unit normal to the instantaneous plane of orbit, while the unit vectors
 $\;{\bf\hat{x}}\,,\,{\bf\hat{y}}\,,\,{\bf\hat{z}}\;$ denote the basis of the
 co-precessing coordinate system $\,\it x\,,\,y\,,\,z\,$. (The axes $\,\it x\,$
 and $\,\it y\,$ belong to the planet's equatorial plane, and the longitude of
 the node, $\,\Omega\,$, is measured from $\,\it x\,$.)

 The quantity $\;\dot{\mu}_{\perp}\;$ is defined as
 \ba
% \nonumber
 \dot{\mu}_{\perp}\;\equiv\;\dot{\mubold}\cdot\wbold\;\;\;,
 \label{44}
 \ea
 but not as $\;d(\mubold\cdot\wbold)/dt\;$ -- a subtlety important
 to our further developments.

Insertion of these expressions into equations (\ref{20} - \ref{21})
will give:
 \ba
 % \nonumber
 \frac{da}{dt}\;=\;-\;2\;a\;\frac{\dot{\mu}_{\perp}}{n}\;\sqrt{1\;-\;e^2}\;=\;-\;
 \frac{2\;a^{5/2}}{\sqrt{G\;m\;}}\;\dot{\mu}_{\perp}\;\sqrt{1\;-\;e^2}~~~,~~~~~~~~~~~~~~
 ~~~~~~~\,
 \label{45}\\
 \nonumber\\
 % \nonumber
 \frac{de}{dt}\;\,=\,\;\frac{5}{2}\;\frac{\dot{\mu}_{\perp}}{n}\;e\;\sqrt{1\;-\;e^2}\;\,
 =\,\;\frac{5}{2}\,\;\frac{a^{3/2}\;\,e}{\sqrt{G\;m\;}}\;\,\dot{\mu}_{\perp}\;\,\sqrt{1
 \;-\;e^2}~~~.~~~~~~~~~~~~~~~~~~~~~
 \label{46}
 \ea
For small eccentricities, the approximate solution is:
 \ba
% \nonumber
 a\;=\;a_o\;\exp\left[\,-\,\frac{\textstyle{2}}{\textstyle{n_o}}\;
 \left(\mu_\perp\;-\;{\mu}_{\perp_o}
 \right)\,\right]^{\textstyle{_{-2/3}}}\;+\;O(e^2)\;\approx\;{a_o}\left[\;1\,+\,
 \frac{\textstyle{4\;\;}}{\textstyle{3\;n_o}}\;
 \left(\mu_\perp\;-\;{\mu}_{\perp_o}
 \right)\;\right]~~~,~\,
 \label{47}
 \ea
 \ba
 e\;=\;e_o\;\exp\left[\,-\,\frac{\textstyle{2}}{\textstyle{n_o}}\;
 \left(\mu_\perp\;-\;{\mu}_{\perp_o}
 \right)\,\right]^{\textstyle{_{-5/4}}}\;+\;O(e^2)\;\approx\;e_o\;\left[\,1\;+\;
 \frac{\textstyle{5\;\;}}{\textstyle{2\;n_o}}\;
 \left(\mu_\perp\;-\;{\mu}_{\perp_o}
 \right)\,\right]~~~,
 \label{48}
 \ea
 where $~n_o~\equiv~\left(G\,m\right)^{1/2}\,a_o^{-3/2}~$. We see that variations in the
 primary's precession exert almost no influence upon the satellite's semimajor axis and
 eccentricity.

 It should, nevertheless, be kept in mind that the satellite orbital elements evolve not
 only under the influence of the primary's precession but also under the action of tides.
 Within the truncated model developed in this paper, we shall neglect the tides, but
 shall introduce them on a subsequent stage of the project.
 ~\\

\subsection{\textbf{The periapse, the inclination, and the node -- in the leading order
 of $~e~$.}}

 Under the assumption of $\;a\;$ and $\;e\;$ remaining virtually unchanged, equations
 (\ref{22} - \ref{24}) will make a closed system, provided we omit the $\;\langle\,
 \mubold\cdot(\,.\,.\,.\,)\,\rangle$ (for the reasons explained above) and also
 substitute the orbital averages of the $\;\dotmubold$-dependent terms with their
 approximations in the leading order of the eccentricity. This level of approximation
 would be consistent with the approximation used in (\ref{47} - \ref{48}). As shown
 in the Appendix,
 \ba
 \langle\;\dotmubold\;\cdot\;\left(\;-\;\efbold\,\times\,\frac{\partial {\efbold}}{
 \partial e}\;\right)\;\rangle\,=\;0~~~,~~~~~~~~~~~~~~~~~~~~~~~~~~~~~~~~~~~~~~~~~~~
 ~~~~~~~~~~~~~~~~~~~~~~~~~~~~~~~~~~
 \label{49}
 \ea
 %~\\
 \ba
 \bra\; \dotmubold\;\cdot
 \left(\;-\;\efbold\;\times\;\frac{\partial \efbold}{\partial
 \omega}\;\right)\;\ket\;=\;-\;a^2\;\left(\;\dot{\mu}_1\;\sin\inc\;\sin\Omega
 \,-\,\dot{\mu}_2\;\sin\inc\;\cos\Omega
 \,+\,\dot{\mu}_3\;\cos\inc
 \;\right)\,+\,O(e^2)~~,~~~
 \label{50}
 \ea
 ~\\
  \ba
 % \nonumber
 \bra\dotmubold\cdot\left(-\efbold\times\frac{\partial{\efbold}}{\partial\Omega}\right)
 \ket=
  % ~~~~~~~~~~~~~~~~~~~~~~~~~~~~~~~~~~~~~~\\
  % \nonumber\\
  % \nonumber\\
 \frac{a^2}{2}\left[-\dot{\mu}_1\,\sin\inc\,\sin\Omega\,\cos\inc+\dot{\mu}_2\,\sin\inc\,
 \cos\Omega\,\cos\inc-\dot{\mu}_3\,\left(2-\,\sin^2\inc\right)\right]+O(e^2)~~,~~
 \label{51}
 \ea
 ~\\
 \ba
 % \nonumber
 \bra\;\dotmubold\cdot\left(\;-\;\efbold\;\times\;
 \frac{\partial {\efbold}}{\partial \inc}\;
 \right)\;\ket\;=\;-\;\frac{a^2}{2}\;\left(\;\dot{\mu}_1\;\cos\Omega\;+
 \;\dot{\mu}_2\;\sin\Omega
 \;\right)\,+\,O(e^2)~~~.~~~~~~~~~~~~~~~~~~~~~~~~~~~~~~~~
 \label{52}
 \ea
Substitution of (\ref{19}) and of the above expressions for the
$\;\dotmubold $-terms into (\ref{22} - \ref{24}) will give us:
 \begin{eqnarray}
 \frac{d\omega}{dt}\,=\,\frac{3\,n\,J_2}{4}\;\,\left(\frac{\rho_e}{a}\right)^2\,\;
 \left(\,5\,\cos^2i\,-\,1\,\right)\,+\,\mu_n\;\cot\inc\,-\,\mu_{\perp}\;+\;
 \frac{1}{2}\;\left(\,\frac{\dot{\mu}_1}{n}\;\cos\Omega\;+\;\frac{\dot{\mu}_2}{n}
 \;\sin\Omega\,\right)\;+\;O(e^2)~~,~~
  \label{53}
  \end{eqnarray}
 ~\\
 \ba
 \frac{d\inc}{dt}\,=\,-\,\mu_1\,\cos\Omega\,-\,\mu_2\,\sin\Omega\,-\,
 \frac{\dot{\mu}_{\perp}}{n}\,\cot
 \inc\,-\,\frac{\dot{\mu}_n}{2n}\,+\,\frac{1}{\sin\inc}\;\frac{\dot{\mu}_3}{n}\;+\;
 O(e^2)\;~~~,~~~~~~~~~~~~~~~~~~~~~~~~~~~~~~~~
 \label{54}
 \ea
  ~\\
  ~\\
  \begin{eqnarray}
 \frac{d\Omega}{dt}\,=\;-\;\frac{3}{2}\;{n\;J_2}\;\left(\frac{\rho_e}{a}\right)^2\;{\cos
 {i}}\;-\,\frac{\;\mu_{n}\,}{\sin\inc}\,+\,
 \frac{1}{2\,\sin\inc}\;\left[\;-\;\left(\,\frac{\dot{\mu}_1}{n}\;\cos\Omega\,+\,
 \frac{\dot{\mu}_2}{n}\;\sin\Omega
 \,\right)\,\right]\;+\;O(e^2)\;\;\;,~~~~~~
 \label{55}
 \ea
 where $\,\mu_{\perp}\,$ and $\,\dot{\mu}_{\perp}\,$ are given by (\ref{43} -
 \ref{44}). The quantity
  \ba
 {{\mu}}_n\;\equiv\;-\;{\mu}_1\;\sin \Omega\;\cos\inc\,+\,
 {\mu}_2\;\cos \Omega\;\cos \inc\,+\,{\mu}_3\;\sin
 \inc\;\;\;,
 \label{56}
 \ea
 is the component of $\;\mubold\;$, pointing from the gravitating centre towards the
 ascending node of the orbit, while
 \ba
 {\dot{\mu}}_n\;=\;-\;{\dot\mu}_1\;\sin \Omega\;\cos\inc\,+\,
 {\dot\mu}_2\;\cos \Omega\;\cos \inc\,+\,{\dot\mu}_3\;\sin
 \inc\;\;\;,
 \label{57}
 \ea
is its time derivative taken in the frame co-precessing with the
satellite orbit plane. (Taking the derivative in this frame, we
differentiate only the components of ${\mubold}$, but not the
angles.)

Under the assumption of constant $\,a\,$ and small $\,e\,$,
equations (\ref{54} - \ref{55}) make a closed system.

 \subsection{Goldreich's approximation}

It would now be tempting to introduce an even stronger assumption
that both $\;|\dotmubold|/\left(\,n^2\,J_2\,\sin\inc\,\right)\;$
and $\;|\mubold|/\left(\,n\,J_2\,\sin\inc\, \right)\;$ are much
less than unity, and to derive therefrom the system
 \begin{eqnarray}
 \frac{d\Omega}{dt}\,\approx~
 -\;\frac{3}{2}\,{n\,J_2}\,\left(\frac{\rho_e}{a}\right)^2
 \;\frac{\cos
 {i}}{\left(1-e^2
 \right)^2}\;\;\;,
 \label{59}\\
 \nonumber\\
% \end{eqnarray}
% \begin{eqnarray}
 \frac{d \inc}{dt}\;\approx\;-\;\mu_1\;\cos \Omega\;-\;
 \mu_2\;\sin \Omega\;\;\;\,{,}~~~~~
 \label{60}
 \end{eqnarray}
whose solution,
 \ba
 \nonumber
 i\,=\;-\,\frac{\mu_1}{\chi}\,\cos \left[\;-\,\chi\,
 \left(t\,-\,t_o\right)\,+\,\Omega_o
 \right]\,+\,\frac{\mu_2}{\chi}\,\sin \left[\;-\,\chi\,
 \left(t\,-\,t_o\right)\,+\,\Omega_o
 \right]\,+\,i_o \;\;\;{,}\\
 \label{61}\\
 \nonumber
 \Omega\,=\,-\,\chi\,\left(t\,-\,t_o\right)\,+\,\Omega_o~~~~~~~
 \mbox{where}\;\;\;~~\chi\;\equiv\;
 \frac{3}{2}\;{n\;J_2}\;\left(\frac{\rho_e}{a}\right)^2\;\;
 \frac{\cos {\inc}}{\left(1\;-\;e^2
 \right)^2}\;\;\;,~
 \ea
 seems to indicate that, in the course of planet precession (the term
 ``precession" including, as agreed above, also nutations and the Chandler
 wobble and polar wander), the satellite inclination oscillates about $\;\inc_o\;$.
 Approximation (\ref{61}) has already appeared in the literature. Goldreich (1965)
 derived such equations for the orbital averages of some nonosculating elements (which
 later were termed, by Brumberg (1992), ``contact elements"). In our case, however, the
 approximation (\ref{61}) was derived for the secular parts of osculating elements.
 We see that, in neglect of $\;\mubold^{\,2}$-terms and under the assumption
 of constant $\;\mubold\;$, the equations for the secular parts of osculating elements
 coincide with those for the secular parts of the contact ones. (For a detailed
 explanation of this fact see Efroimsky (2005a).)

 The evident flaw of approximation (\ref{59} - \ref{61}) is its invalidity in
 the closest vicinity of the equator. In this vicinity, the parameters
$\;|\dotmubold|/\left(\,n^2\,J_2\,\sin\inc\,\right)\;$ and
$\;|\mubold|/\left(\,n\,J_2\,\sin\inc\, \right)\;$ are no longer small; so the
entire approximation falls apart and gives no immediate indication on whether the
inclination will go through zero and alter its sign or will ``bounce off" the
equator. At the first glance, this technical subtlety does not affect this
approximation's main physical outcome, one that the inclination remains limited and
shows no secular increase. In reality, though, the matter needs further
exploration. For example, if the orbit keeps bouncing off the equator and the sign
of $\;\inc\;$ stays unaltered for long, then the term $\;\dot{\mu}_n/(2n)\;$    in
(\ref{54}) may, potentially, keep accumulating through aeons, creating a drift of
the inclination. Whether this is so or not can be learned numerically through a
more accurate approximation based on equations (\ref{53} - \ref{55}). A more
definite thing is that the Goldreich approximation is intended only for low
inclinations: as can be seen from equation (\ref{55}), at high inclinations it will
fail, because the term $\;\mu_{\perp}/\sin\inc\;$ will dominate over the
$\;J_2\,\cos\inc\,$ term. All these issues will be addressed in our subsequent
paper (Gurfil, Lainey \& Efroimsky 2007).

 \subsection{Can precession cause secular changes of the inclination?}

 Above we saw that the Goldreich approximation reveals no secular terms
 in the expression for the inclination relative to the moving equator.
 While a reliable quest into this matter will demand numerical
 integration of the entire system of the planetary equations, we shall
 try to work out a qualitative estimate based on the approximation
 less crude than that of Goldreich. To this end we shall plug (\ref{44})
 and (\ref{56}) into (\ref{54}), and shall omit all the long-period
 terms. Thus we shall be left with the following estimate for the
 secular part of $\,\inc\,$:
 \ba
 \frac{d\inc^{^{(sec)}}}{dt}\;=\;\frac{\dot{\mu}_3}{2\,n}\;\sin\inc\;+\;
 O(e^2)\;+\;\mbox{long-period~terms}\;\;\;
 \label{equation}
 \ea

 \subsubsection{Small initial inclinations}

 For small inclinations, the above equation will look:
 \ba
 \frac{d\inc^{^{(sec)}}}{dt}\;\approx\;A\;\inc^{^{(sec)}}\;\;\;
 \label{}
 \ea
 where
 \ba
 A\;\equiv\;\frac{\dot{\mu}_3}{2\,n}\;\approx\;\frac{\ddot{h}_p\;\cos
 I_p}{2\,n}\;\approx\;\frac{\ddot{h}_p}{2\,n}\;\;\;,
 \label{}
 \ea
 $h_p\;$ and $\;I_p\;$ being the longitude of the node and the inclination of
 the equator of date on that of epoch. (see Appendix A8). From here we see that
 the osculating component of $\;\inc\;$ will, approximately, obey
 \ba
 \inc^{^{(sec)}}\;\approx\;\inc_o\;e^{At}\;\;\;.
 \label{secular}
 \ea
 The exponential dependence evidences of the presence of chaos in the system. It
 should be mentioned, though that the chaos will be weak, because $\,A\,$ is
 extremely small. Besides, the second derivative of the precessing equator's node,
 $\,\ddot{h}_p\,$, which enters the expression for $\,A\,$, does not keep the same
 sign through aeons. The rate, at which node $\,h_p\,$ and the inclination $\,I_p
 \,$ evolve, can be computed, via the Colombo equation, from the rate of precession
 of the planet's orbit about the Sun. (For details on the Colombo model see the
 Appendix below.) Qualitatively, one may expect the spectrum of $\,h_p\,$ and
 $\,I_p\,$ to resemble the frequency content of the planet's orbital precession.
 (See the table in the Appendix.) On all these grounds, the time dependence of
 $\,\inc\,$ is constituted by the high-frequency oscillations (\ref{61})
 superimposed on a much slower evolution (\ref{secular}). We expect this slow
 evolution to look as a saw-tooth plot, because the sign of $\,\ddot{h}_p\,$
 (and therefore of $\,A\,$) alters from time to time for the reason explained
 above. Due to this saw-tooth nature of the long-term evolution of $\,\inc^{(sec)}
 \,$, no considerable secular increase of the satellite's inclination should be
 expected, at least in the case of a small initial $\,\inc_o\,$. Numerical
 calculations performed in the $\;e^3\;$ order confirm this conclusion. Moreover,
 it turns out that even at not too small initial inclinations no secular changes in
 $\,\inc\,$ accumulate over time scales of order billion years.

 The said numerical results and plots are presented in our subsequent paper
 Gurful, Lainey \& Efroimsky (2007), devoted to a numerical implementation
 of our semianalytical model in the $\;e^3\;$ order.

  \subsubsection{Large initial inclinations}

 For near-polar orbits, the equation (\ref{equation}) will read:
 \ba
 \frac{d\inc^{^{(sec)}}}{dt}\;\approx\;A\;\;\;,
 \label{}
 \ea
 whence
 \ba
 \inc^{^{(sec)}}\;\approx\;A\;t\;\;\;.
 \label{}
 \ea
 Once again, due to the undulatory sign alterations of $\,A\,$, we shall get a
 ``saw-tooth" behaviour, though this time the teeth will be less steep than in
 the small-inclination case governed by the exponent (\ref{secular}). The teeth
 will be expected to cross the polar orbit once in a while. This kind of time
 dependence (so-called ``crankshaft") is indeed what results from the numerical
 computations. (Ibid.)

 All in all, unless we begin very close to the pole, the variable equinoctial
 precession is not expected to entail secular changes in the satellite inclination
 relative to the equator of date. Exceptional is the case of near-polar orbits:
 in that case, leaps across the pole are possible. (See Ibid. for details and plots.)

 \section{Preparation for computation in the $\;e^3\;$ order}

% Each and every model is somehow truncated. Our one does not yet incorporate the
% tidal or
% relativist effects, nor takes it into account the mutual attraction of satellites or
% the planet's triaxiality. Nonetheless, we shall take into account terms of orders $\,
% e^2\,$ and $\,e^3\,$, because in realistic situations they may be much larger than the
% afore mentioned omitted inputs.

 Insertion of (\ref{19}) into equations\footnote{~As within this model both the
 Hamiltonian perturbation and the $\dotmubold$-dependent terms are substituted
 with their orbital averages, $\,M_o\,$ becomes a nuisance parameter,
 so the planetary equation for $\,dM_o/dt\,$ is omitted.} (\ref{20} - \ref{24})
 will lead us to the following system:
 \ba
 \frac{da}{dt}\,=\;-\;2\;\frac{\dot{\mu}_{\perp}}{n}\;\,a\;\,\left(1\,-\,e^2\right)^{1/2}
 \;\;\;,\;\;~~~~~~~~~~~~~~~~~~~~~~~~~~~~~~~~~~~~~~~~~~~~~~
 \label{63}
  \ea
  \begin{eqnarray}
 \frac{de}{dt}\,=\;\frac{5}{2}\;\frac{\dot{\mu}_{\perp}}{n}\;e
 \;\left(\,1\;-\;e^2\,\right)^{1/2}~~~,~~~~~~~~~~~~~~~~~~~~~~~~~~~~~~~~~~~~~~~~~~~~~~~~
 \label{}
  \end{eqnarray}
  ~\\
  \begin{eqnarray}
 \frac{d\omega}{dt}=\frac{3}{2}\frac{n\,J_2}{\left(1-e^2\right)^2}\left(
 \frac{\rho_e}{a}\right)^2\left(\frac{5}{2}\,\cos^2\inc-\frac{1}{2}\right)-
 {\mu_{\perp}}+{\mu}_n\,\cot\inc\,
% \ea
% ~\\
% \ba
 - \frac{\cos\inc}{n a^2(1-e^2)^{1/2} \sin \inc }\;\langle\dotmubold\left(
 -\efbold\times\frac{\partial \efbold}{\partial \inc} \right)\rangle
 \;,~~~
  \end{eqnarray}
  ~\\
 \ba
 \nonumber
 \frac{d \inc}{dt}\;=~-\,\mu_1\,\cos\Omega\,-\,\mu_2\,\sin\Omega\,~~~~~~~~~~~~~~~~~~
 ~~~~~~~~~~~~~~~~~~~~~~~~~~~~~~~~~~~~~~~~~~~~~~~~~~~~~~~~~~~~~~~
 \ea
 \ba
 ~~~~\,+\,\frac{\cos \inc}{n a^2(1 - e^2)^{1/2} \sin
 \inc}\;
 \langle \;\dotmubold\left(\,-\,\efbold\times
 \frac{\partial \efbold}{\partial \omega}  \right)\,
 \rangle\;-\,\frac{1}{na^2\,(1-e^2)^{1/2}\,\sin \inc
 }\;\langle \, \dotmubold\, \left(\,-\,\efbold\,\times\,
 \frac{\partial \efbold}{\partial \Omega}  \right)\;
 \rangle\;\;\;{,}\;\;\;\;~~~
  \ea
 ~\\
 ~\\
 \ba
 \frac{d\Omega}{dt}\,=\,-\,\frac{3}{2}\,n\,J_2\,\left(\,\frac{\rho_e}{a}\,\right)^2\;
 \frac{\cos\inc}{\left(1\;-\;e^2\right)^2}\;-\;\frac{\mu_n}{\sin \inc}\;
 +\;\frac{1}{n\; a^2\;(1\;-\;e^2)^{1/2}\;\sin \inc }\;\langle\,\dotmubold\left(
 \,-\,\efbold\times\frac{\partial\efbold}{\partial\inc}\right)\,\rangle\,\;\;,\;\;\;\;~~~
 \label{67}
  \ea
 % ~\\
 % ~\\
 % \begin{eqnarray}
 % \nonumber
 %\frac{d M_o}{dt}\,=\,\;~~~~~~~~~~~~~~~~~~~~~~~~~~~~~~~~~~~~~~~~~~~~~~~~~~~~~~~~~~~~
 % ~~~~~~~~~~~~~~~~
 % \ea
 % \ba
 % \nonumber
 % -\,\;\frac{1\,-\,e^2}{n\,a^2\,e}\,\; \left[\;\frac{\partial \left(\,-\,\Delta
 % {\cal H}^{(eff)} \right)}{\partial e }~ + ~\langle~\mubold\cdot
 % \left(\frac{\partial{\efbold}}{\partial e}\times {\bf{\vec g}}\,-\,{\efbold}\times
 % \frac{\partial{\vec{\bf g}}}{\partial e}\right)~\rangle \;-\;\langle \; \dotmubold\;
 % \left(\efbold\,\times\,\frac{\partial \efbold}{\partial e}\right)\;\rangle\;\right]\\
 % \nonumber\\
 % \label{68}\\
 % \nonumber\\
 % \nonumber
 % -\;~\frac{2}{n\,a}\,\left[\;\frac{\partial \left(\,-\,\Delta {\cal H}^{(eff)} \right)
 % }{\partial a }~ + ~ \langle ~\mubold\cdot\left(\frac{\partial{\efbold}}{\partial a}
 % \times {\bf{\vec g}}\,-\,{\efbold}\times\frac{\partial{\vec{\bf g}}}{\partial a}
 % \right)~\rangle\;-\;\langle \; \dotmubold\; \left(\efbold\,\times\,\frac{\partial
 % \efbold}{\partial a} \right)\;\rangle\; \right] \;\;,~~~~~
 % \end{eqnarray}
  ~\\
 where, according to the Appendix,
 \ba
 \nonumber
 \bra\;\dotmubold\cdot\left(\;-\;\efbold\;\times\;
 \frac{\partial {\efbold}}{\partial \inc}\;
 \right)\;\ket\;=~~~~~~~~~~~~~~~~~~~~~~~~~~~~~~~~~~~~~~~~~~~~~~~~~~~~~~~~~~~~~~~~~~~
 ~~~~~~~~\\
 \nonumber\\
 \nonumber\\
 \nonumber
 \frac{a^2}{4}\;\left\{\;
 \dot{\mu}_1\;\left[\;-\;\left(2\,+\,3\,e^2\right)
 \,\cos\Omega\,+\,5\,e^2\;\left(\cos\Omega\;\cos 2\omega\,-\,\sin
 \Omega\;\sin 2\omega\;\cos\inc \right)
 \;\right]\;+ \right.\\
 \nonumber\\
 \nonumber\\
 \nonumber
 \dot{\mu}_2\;\left[\;-\;\left(2\,+\,3\,e^2\right)
 \,\sin\Omega\,+\,5\,e^2\;\left(\sin\Omega\;\cos 2\omega\,+\,\cos
 \Omega\;\sin 2\omega\;\cos\inc \right)
 \;\right]\;+\,\\
 \nonumber\\
 \nonumber\\
% \nonumber
 \left. \dot{\mu}_3\;\left[\;5\;e^2\;\sin
 2\omega\;\sin\inc
 \;\right]\;\right\}~~~~~~~~~~~~~~~~~~~~~~~~~~~~~~~~~~~~~~~~~~~~~~~~~~~~~~~~~~~~~~\,
 \label{69}
 \ea
  \ba
 \bra\;\dotmubold\,\cdot\,
 \left(\,-\,\efbold\;\times\;\frac{\partial \efbold}{\partial
 \omega}\;\right)\;\ket\,=\,
 -\;\frac{a^2}{2}\;
 \left(2\;+\;3\;e^2\right)\;\left(\;\dot{\mu}_1\;\sin\inc\;\sin\Omega
 \;-\;\dot{\mu}_2\;\sin\inc\;\cos\Omega
 \;+\;\dot{\mu}_3\;\cos\inc
 \;\right)~~~,~~~~
 \label{71}
 \ea
 \ba
 \nonumber
 \bra\;\dotmubold\cdot\left(\;-\;\efbold\;\times\;
 \frac{\partial {\efbold}}{\partial \Omega}\;
 \right)\;\ket\;=
 ~~~~~~~~~~~~~~~~~~~~~~~~~~~~~~~~~~~~~~~~~~~~~~~~~~~~~~~~~~~~~~~~~~~~~~~~~~~~~~~~~
 ~~~~~~~~~~~\\
 \nonumber\\
 \nonumber\\
 \nonumber
 \frac{a^2}{4}\;\left\{\;
 \dot{\mu}_1\;\sin\inc\;
\left[\;-\;\left(2\,+\,3\,e^2\right)\;\sin\Omega\;\cos\inc\;+\;5\;e^2\;\left(\,
\cos \Omega\;\sin 2\omega\,+\,\sin\Omega\;\cos 2\omega\;\cos
\inc \,\right) \;\right] \right. \\
 \nonumber\\
 \nonumber\\
 \nonumber
 +\;\dot{\mu}_2\;\sin\inc\;\left[\;
 \left(2\,+\,3\,e^2\right)\;\cos\Omega\;\cos\inc\;+\;5\,e^2\;\left(
 \sin\Omega\;\sin 2\omega\;-\;\cos\Omega\;\cos 2\omega\;\cos\inc
 \right)
 \;\right]~~~~\\
 \nonumber\\
 \nonumber\\
 \left. \;-\;\dot{\mu}_3\;\left[\;
 \left(2\,+\,3\,e^2\right)\,\left(2\;-\;\sin^2\inc\;\right)\;+\;
 {5}\,e^2\,\sin^2\inc\;\cos 2\omega
 \;\right]
 \;\right\}
 ~~~~~~~~~~~~~~~~~~~~~~~~~~~~~~~~~~~
 \label{73}
 \ea

\section{Conclusions}

 In this article we continued our analytical investigation of the behaviour of orbits
 about a precessing oblate planet. We built up a reasonably simplified model that takes
 into account both the long-term variability of the planetary precession (variability
 caused by the planet's orbit precession) and the short-term variability (polar wonder,
 etc).\\
 ~\\
 We have written down equations (\ref{63} - \ref{67}) that describe evolution of the
 satellite orbit at long time scales. These equations also include known functions of
 time, $\;\mu_1\,,\;\mu_2\,,\;\mu_3\,,\;\mu_{\perp}\,,\;\mu_n\,,\;\dot{\mu}_1\,$, $\;
 \dot{\mu}_2\,,\;\dot{\mu}_3\,,\;\dot{\mu}_{\perp}\,,\;\dot{\mu}_n\;$, which are
 various projections of the planet axis' precession rate. An algorithm for computation
 of these functions of time is presented in the Appendix. These functions vary in time
 as a result of precession of the primary's orbit about the Sun. This way, we have
 analytically established connection between the precession of the planet's orbit and
 the evolution of its satellites. Physically, this connection comes into being through
 the following concatenation of circumstances: precession of the planetary orbit leads
 to variations in the Solar torque acting on the planet; the torque variations cause
 changes in the planet axis' precession; these changes, in their turn, entail
 variations of orbits of the planet's satellite. This effect is extremely weak and
 accumulates over very long time scales.\\
~\\
 \noindent All in all, we have fully prepared a launching pad for computation of the
 evolution of near-equatorial circummartian orbits at long time scales. The methods
 and results of this integration, and their physical interpretation will be presented
 in our next publication (Gurfil, Lainey, \& Efroimsky 2007), which will also include
 the pull of the Sun. Briefly speaking, those results are to be three-fold. First, it
 turns out that our semianalytical model is robust beyond expectations. Despite the
 averaging and the neglect of $\,\mubold^{{{\,2}}}$-terms, it works very well over
 timescales up to, at least, 20 Myr. Second, it turns out that precession by itself
 (i.e., in the presence of the Sun but in the absence of the other physical factors
 like the tides and the planet's triaxiality) cannot cause accumulating secular
 changes in
 the satellite inclination, {\emph{provided the initial inclination is not too
 large}}. This means that, for orbits not too close to the polar one, the main
 prediction of the Goldreich model stays valid, even though the model cannot
 adequately describe the entire dynamics (which becomes weakly chaotic). Third,
 it turns out that in the vicinity of the polar orbit precession of
 the primary can cause major alteration of the satellite orbits, including unusual
 features in the behaviour of the inclination. See {\emph{Ibid}}
 for more details. Further work along this line of research will be aimed at
 including more factors into the model -- the tidal forces, the pull of the Sun,
 the triaxiality of the planet, etc.\\
~\\

{\underline{\bf{\Large{Acknowledgments}}}}\\
~\\
 I am indebted to Pini Gurfil, George Kaplan, and Valery Lainey, with whom I had
 numerous fruitful discussions on the topic addressed in this article. My profoundest
 gratitude goes to Victor Slabinski who kindly read the manuscript and made comments
 critical for its quality. This research was supported by NASA grant W-19948.
 ~\\

 \noindent
{\underline{\bf{\Large{Appendix.}}}} \\
 ~\\
The goal of this Appendix is to calculate, in neglect of
nutation-caused resonances, the secular parts of the
$\,\mubold$-dependent terms emerging in the planetary equations
(\ref{19}) -- (\ref{25}). We shall also explain how to compute the
time dependence of various projections of the planetary precession
rate $\,\mubold\,$ and of its time derivatives.

 %  \textbf{Each section of the Appendix below will be a
 %  continuation of the appropriately numbered section
 %  from the Appendix in Efroimsky (2004).}
 ~\\

\noindent {{\bf{\large{A.1.~~~The averaging rule\\}}}}

 The mean values are to be calculated via the averaging rule:
 \ba
 \langle\;\,.\,.\,.\,\;\rangle\;\equiv\;\frac{\left(1\;-\;e^2\right)^{3/2}}{2\;\pi}\;
 \int_{-\pi}^{\pi}\;.\,.\,.\;\;\;\frac{d\nu}{\left(1\;+\;e\;\cos \nu  \right)^2}
 \label{A1}
 \label{rule}
 \ea
 Since the averaging is carried out over the true anomaly, it will
 be convenient to express the precession rate not as a function of
 time, $\;\mubold(t)\;$, but as a function of the true anomaly:
 \ba
 \mubold(\nu)\;=\;\int_{0}^{\infty}\;\left[
 \;\mubold^{(s)}(W)\;\sin({W\nu})\;+\;
 \mubold^{(c)}(W)\;\cos({W\nu})\;\right]\;dW\;\;\;.
 \label{A2}
 \ea
$W\,$ being the circular ``frequency" related to the true
 anomaly $\;\nu\;$.

In what follows, we shall need the average of the projection of
$\;\mubold(t)\;$ onto the instantaneous normal to the orbit. This
projection, $\,\mu_{\perp}\,$, will be expressed by
 \ba
 \mu_{\perp}\;\equiv\;\mubold\,\cdot\,\wbold\;=\;\mu_1\;\sin \inc\;\sin \Omega\;-
 \;\mu_2\;\sin \inc\;\cos \Omega\;+\;\mu_3\;\cos \inc\;\;\;\;.
 \label{A4}
 \ea
where the unit vector
 \ba
 {\wbold}\;=\;{\bf\hat{x}}\;\sin \inc\;\sin \Omega\;-
 \;{\bf\hat{y}}\;\sin
 \inc\;\cos \Omega\;+\;{\bf\hat{z}}\;\cos
 \inc\;\;~~~~~~~~~~~~~~~~~~~
 \label{A5}
 \ea
stands for the normal to the instantaneous osculating ellipse, and
the unit vectors
$\;{\bf\hat{x}}\,,\,{\bf\hat{y}}\,,\,{\bf\hat{z}}\;$ are the basis
of the co-precessing coordinate system $\,\it x\,,\,y\,,\,z\,$. (The
axes $\,\it x\,$ and $\,\it y\,$ belong to the planet's equatorial
plane, and the longitude of the node, $\,\Omega\,$, is measured from
$\,\it x\,$.) Expressions of $\;\mu_j\;$ via the longitude of the
node and inclination of the equator of date relative to that of
epoch are given below in section {{A.12}}.

 \ba
 \nonumber
 \mbox{{\bf{\large{ \textbf{A.2.~~~Calculation of the secular part of}}}}}
 \;\;\;\;\dotmubold\cdot\left(\;-\;\efbold\;\times\;
 \frac{\partial {\efbold}}{\partial a}\;
 \right)\;\;\;.~~~~~~~~~~~~~~~~~~~~~~~~~
 ~~~~~~~~~~~~~~~~~~~~~~~~~~
 \ea

According to the Appendix to Efroimsky (2004), the term of our
concern, written in terms of the orbital elements, is given by
 \ba
  \nonumber
 \dotmubold\;\cdot\;
 \left(\;-\;\efbold\,\times\,\frac{\partial {\efbold}}{\partial
 a}\,\right)\;=\;\frac{3}{2}\,\;\dot{\mu}_{\perp}\;\;a\;n\;\left(\;t\;-\;t_o\;\right)
 \;\sqrt{1\;-\;e^{2}}\;\;\;,
 \label{A75}
 \ea
expression linear in $\;t\;-\;t_o\;$. It coincides with its
secular part, if we assume that the spectrum of $\;\mubold\;$
lacks short-period modes.

% To proceed, we shall need the following formula:
% \ba
% \frac{\partial t}{\partial \nu}\;=\;\frac{\partial t}{\partial
% E}\;\;\frac{\partial E}{\partial \nu}\;=\;\frac{1}{n}\;\frac{1\;-\;e^{2}}{1\;+\;
% e\;\cos \nu}\;\frac{\sqrt{1\;-\;e^2}}{1\;+\;e\;\cos\nu}\;=\;
% \frac{1}{n}\;\frac{\left(1\;-\;e^{2}\right)^{3/2}}{\left(1\;+\;
% e\;\cos \nu\right)^2}\;\;\;.
% \label{A771}
% \ea
% (Here the expression for $\;\partial E/\partial \nu\;$ is borrowed
% from (\ref{A-2}), while the formula for $\;\partial E/\partial t \;$
% can be derived from the Kepler equation and the first
% equation of (\ref{A-1}).) Insertion of (\ref{A771}) and (\ref{45})
% into (\ref{A70}) results in
%  \ba
%  -\;\efbold\;\times\;\frac{\partial {\efbold}}{\partial a}\;=\; -\;
% \sqrt{G\,m\,a\,\left(1\;-\;e^2\right)}\;\;\;{\bf\vec w} \;\;\;
% \frac{1}{n}\;\;\frac{\left(1\;-\;e^{2}\right)^{3/2}}{\left(1\;+\;
% e\;\cos \nu\right)^2}\;\;\left(\;\frac{\partial \nu}{\partial
% a}\;\right)_{t,\,e,\,M_o}\;\;\;.
% \label{A772}
% \ea

 \ba
 \nonumber
 \mbox{{\bf{\large{\textbf{A.3.~~~Calculation of the secular part of}}}}}
 \;\;\;\;\dotmubold\cdot\left(\;-\;\efbold\;\times\;
 \frac{\partial {\efbold}}{\partial e}\;
 \right)\;\;\;.~~~~~~~~~~~~~~~~~~~~~~~~~
 ~~~~~~~~~~~~~~~~~~~~~~~~~~
 \ea

 From Efroimsky (2004) we have:
  \ba
 \dotmubold\;\cdot\;\left(\;-\;\efbold\,\times\,\frac{\partial {\efbold}}{\partial
 e}\;\right)\,=\;
-\;\dot{\mu}_{\perp}\;\;a^2
 \;\;\frac{\left(1-e^{2}\right)}{1+
 e\;\cos \nu}\;\sin \nu\;\;\;.\;\;\;\;
 \;\;\;.\;
 \label{A80}
 \ea
The secular part of this expression, calculated through
  \ba
 \langle\;\dotmubold\;\cdot\;\left(\;-\;\efbold\,\times\,\frac{\partial
 {\efbold}}{\partial e}\;\right)\;\rangle\;=\;
-\;\dot{\mu}_{\perp}\;\;a^2\;\;\left(1-e^{2}\right)
 \;\;\frac{\left(1-e^{2}\right)^{3/2}}{2\;\pi}\;\;\int^{\pi}_{-\pi}\;
 \frac{\sin\nu\;\;d\nu}{\left(1+
 e\;\cos \nu\right)^3}\;\;\;\;,\;\;\;\;
 \;\;\;.\;
 \label{}
 \ea
will vanish, because the function in the numerator is odd, while
the expression in the denominator is even.

 \ba
 \nonumber
 \mbox{{\bf{\large{\textbf{A.4~~~Calculation of the secular and long-period
 parts
 of}}}}}
 \;\;\dotmubold\cdot\left(\;-\;\efbold\;\times\;
 \frac{\partial {\efbold}}{\partial \omega}\;
 \right)\;\;.~
 \ea

We have from Efroimsky (2004):
 \ba
%
% THIS EXPRESSION HAS UNDERGONE
%
% A COMPREHENSIVE CHECK
%
 \dotmubold\;\cdot\;\left(\;-\;\efbold\;\times\;\frac{\partial \efbold}{\partial
 \omega}\;\right)\;=\;-\;\dot{\mu}_{\perp}\;\;a^2\;\;\frac{\left(1\;-\;e^2\right)^2
 }{\left(1\;+\;e\;\cos\nu\right)^2}\;\;\;\;.~~~~~~~~
 \label{briefly}
 \ea
This expression has the following secular part:
 \ba
 \bra\; \dotmubold\;\cdot\;
 \left(\;-\;\efbold\;\times\;\frac{\partial \efbold}{\partial
 \omega}\;\right)\;\ket\;=\;
 -\;\dot{\mu}_{\perp}\;a^2\;\left(1-e^2\right)^2\;\frac{\left(1-e^2\right)^{3/2}}{2
 \;\pi}\;\int_{-\pi}^{\pi} \,\frac{d\nu}{\left(1\;+\;e\;\cos \nu\right)^4}\;=~~~~~
 \ea
 ~\\
 \ba
 \nonumber
-\;\dot{\mu}_{\perp}\;a^2\;\frac{\left(1-e^2\right)^{7/2}}{2\;\pi}\;
\pi\;\frac{2\;+\;3\;e^2}{\left(1\,-\,e^2
 \right)^{7/2}}~=\;-\;\dot{\mu}_{\perp}\;\frac{a^2}{2}\;\left(2\;+\;3\;e^2
 \right)\;=\\
 \nonumber\\
 \nonumber\\
 \nonumber
 -\;\frac{a^2}{2}\;
 \left(2\;+\;3\;e^2\right)\;\left(\;\dot{\mu}_1\;\sin\inc\;\sin\Omega
 \;-\;\dot{\mu}_2\;\sin\inc\;\cos\Omega
 \;+\;\dot{\mu}_3\;\cos\inc
 \;\right)
 \ea
Be mindful that the time derivative of $\;{\mu}_{\perp}\;$,
calculated with aid of (\ref{A4}), reads as
 \ba
 \dot{\mu}_{\perp}\;=\;\dot{\mu}_1\;\sin\inc\;\sin\Omega
 \;-\;\dot{\mu}_2\;\sin\inc\;\cos\Omega\;+\;\dot{\mu}_3\;\cos\inc
 \label{A3}
 \ea
and contains only derivatives of $\;\mu_j\;$ but not of the angles.
This happens because $\;\dot{\mu}_{\perp}\;$ is not a full time
derivative but is defined through
$\;\dot{\mu}_{\perp}\;\equiv\;\dotmubold\;\cdot\;{\wbold}\;$.

 \ba
 \nonumber
 \mbox{{\bf{\large{\textbf{A.5.~~~Calculation of the secular and
 long-period parts of}}}}}
 \;\;\dotmubold\cdot\left(\;-\;\efbold\;\times\;
 \frac{\partial {\efbold}}{\partial \Omega}\;
 \right)\;\;.~
 \ea

In this subsection and below we shall need the following auxiliary
integrals:
%
% THIS EXPRESSION HAS UNDERGONE
%
% A COMPREHENSIVE CHECK
%
 \ba
 \Upsilon_0\;\equiv\;\int_{-\pi}^{\pi}
 \frac{1}{\left(1\,+\,e\,\cos\nu
 \right)^4}\;d\nu\;=\;\pi\;\frac{2\;+\;3\;e^2}{\left(1\,-\,e^2
 \right)^{7/2}}~~~,~~~~~~~~~~~~~~~~~~~~~
 \label{J0}
 \ea
%
% THIS EXPRESSION HAS UNDERGONE
%
% A COMPREHENSIVE CHECK
%
 \ba
 \Upsilon_1\;\equiv\;\int_{-\pi}^{\pi}
 \frac{\sin(\omega +\nu)\;\cos(\omega +\nu)}{\left(1\,+\,e\,\cos\nu
 \right)^4}\;d\nu\;=\;\frac{5}{2}\;\pi\;e^2\;\frac{\sin (2\omega)}{\left(1\,-\,e^2
 \right)^{7/2}}~~~,~~~~~
 \label{J1}
 \ea
%
% THIS EXPRESSION HAS UNDERGONE
%
% A COMPREHENSIVE CHECK
%
 \ba
  \Upsilon_2\;\equiv\;\int_{-\pi}^{\pi}
 \frac{\sin^2(\omega +\nu)}{\left(1\,+\,e\,\cos\nu
 \right)^4}\;d\nu\;=\;\frac{1}{2}\;\pi\;\frac{2\,+\,3\,e^2\,-\,5\,e^2\,
 \cos (2\omega)}{\left(1\,-\,e^2
 \right)^{7/2}}~~.
 \label{J2}
 \ea

\underline{The first component:}\\
 \ba
 %
 % THIS EXPRESSION HAS UNDERGONE A COMPREHENSIVE CHECK
 %
 \nonumber
\bra\; \dot{\mu}_1\;\left(\;\frac{\partial \efbold}{\partial
\Omega}\;\times\;\efbold\;\right)_1\;\ket\;=\;
 ~~~~~~~~~~~~~~~~~~~~~~~~~~~~~~~~~~~~~~~~~~~~~~~~~~~~~~~~~~~~~~~~~~~~~~~~~~~~~~~\\
 \label{A84}\\
  \nonumber
\bra\;\dot{\mu}_1\;a^2\;\;\frac{\left(1\;-\;e^2\right)^2}{\left(1\;+\;e\;\cos
\nu\right)^2}\;\;\left[\;
           \cos\Omega\;\cos(\omega +\nu)\;-\;
           \sin \Omega\;\sin(\omega +\nu)\;\cos \inc
  \;\right]\;\sin(\omega + \nu)\;\sin \inc\;\ket\\
 \nonumber\\
 \nonumber\\
 \nonumber
 =\;\dot{\mu}_1\;a^2\;\left(1\,-\,e^2  \right)^{2}\;\frac{\left(1\,-\,e^2
 \right)^{3/2}}{2\;\pi}\;\left\{\;\Upsilon_1\;\cos\Omega\;\sin \inc\;-\;
 \Upsilon_2\;\sin\Omega\;\cos\inc\;\sin\inc
 \;\right\}
 \ea
 \ba
  \nonumber
=\;\dot{\mu}_1\;\frac{a^2}{4}\;\sin
\inc\;\left\{\;-\;\left(2\,+\,3\,e^2\right)\;\sin\Omega\;\cos\inc\;+\;5\;e^2\;\left[\,
\cos \Omega\;\sin(2\omega)\,+\,\sin\Omega\;\cos(2\omega)\;\cos
\inc \,\right] \;\right\}
 \ea
~\\
~\\
\underline{The second component:}\\
 \ba
 %
 % THIS EXPRESSION HAS UNDERGONE A COMPREHENSIVE CHECK
 %
 \nonumber
 \bra\;\dot{\mu}_2\;\left(\;\frac{\partial \efbold}{\partial \Omega}\;\times\;\efbold
 \;\right)_2\;\ket\;=\;
 ~~~~~~~~~~~~~~~~~~~~~~~~~~~~~~~~~~~~~~~~~~~~~~~~~~~~~~~~~~~~~~~~~~~~~~~~~~~~~~\\
 \label{A85}\\
 \nonumber\\
 \nonumber
\bra\;\dot{\mu}_2\;a^2\;\;\frac{\left(1\;-\;e^2\right)^2}{\left(1\;+\;e\;\cos
\nu\right)^2}\;\;\left[\;
                 \sin\Omega\;\cos(\omega +\nu)\;+\;
                \cos \Omega\;\sin(\omega +\nu)\;\cos \inc
                   \;\right]\;\sin (\omega + \nu)\;\sin
                   \inc\;\ket\\
 \nonumber\\
 \nonumber\\
 \nonumber
=\;\dot{\mu}_2\;a^2\;\left(1\,-\,e^2
\right)^2\;\frac{\left(1\,-\,e^2
\right)^{3/2}}{2\;\pi}\;\left\{\;\Upsilon_1\;\sin\Omega\;\sin\inc\;+\;
\Upsilon_2\;\cos\Omega\;\sin\inc\;\cos\inc\;\right\}\;=
 \ea
 \ba
 \nonumber
\dot{\mu}_2\,a^2\,\frac{\left(1\,-\,e^2
\right)^{7/2}}{2\;\pi}\,\left\{\;\left[\,\frac{5}{2}\;\pi\;e^2\;
\frac{\sin(2\omega)}{\left(1\,-\,e^2\right)^{7/2}}
\right]\,\sin\Omega\,+\, \left[
\frac{1}{2}\;\pi\;e^2\;\frac{2\,+\,3\,e^2\,-\,5\,e^2\,
 \cos (2\omega)}{\left(1\,-\,e^2
 \right)^{7/2}}
\right]\,\cos\Omega\,\right\}\;\sin\inc
 \ea
 \ba
 \nonumber
 =\;\dot{\mu}_2\;\frac{a^2}{4}\;\sin\inc\;\left\{\;
 \left(2\,+\,3\,e^2\right)\;\cos\Omega\;\cos\inc\;+\;5\,e^2\;\left[
 \sin\Omega\;\sin(2\omega)\;-\;\cos\Omega\;\cos(2\omega)\;\cos\inc
 \right]
 \;\right\}
 \ea
~\\
~\\
\underline{The third component:}\\
 \ba
 %
 % THIS EXPRESSION HAS UNDERGONE A COMPREHENSIVE CHECK
 %
\bra\;\dot{\mu}_3\;\left(\;\frac{\partial \efbold}{\partial
\Omega}\;\times\;\efbold\;\right)_3\;\ket
 \;=\;\bra\;
 -\;\dot{\mu}_3\;a^2\;\frac{\left(1 - e^2\right)^2}{\left(1 + e\;\cos
 \nu\right)^2}\;\left[\,\cos^2(\omega + \nu)\,+\,\sin^2(\omega +
 \nu)\;\cos^2\inc \,\right]\;\ket ~=~~~~~
 \label{A86}
 \ea
 \ba
 \nonumber
 -\dot{\mu}_3\,a^2\,
 \frac{\left(1-e^2 \right)^{7/2}}{2\;\pi}\left(\Upsilon_0-\Upsilon_2\sin^2\inc
 \right)=\,-\dot{\mu}_3\,a^2\,\frac{\left(1 - e^2
 \right)^{7/2}}{2\;\pi}\left\{\pi\,\frac{2 + 3e^2}{\left(1-e^2
 \right)^{7/2}} - \pi \frac{2+3e^2-5e^2\,
 \cos (2\omega)}{2\,\left(1\,-\,e^2
 \right)^{7/2}} \,\sin^2\inc\right\}
 \ea
 \ba
 \nonumber
 =\;-\;\dot{\mu}_3\;\frac{a^2}{4}\;\left\{\;
 \left(2\,+\,3\,e^2\right)\,\left[2\;-\;\sin^2\inc\;\right]\;+\;
 {5}\,e^2\,\sin^2\inc\;\cos(2\omega)
 \;\right\}
 \ea

~\\
~\\
\underline{Total:}\\
 \ba
 \nonumber
 \bra\;\dotmubold\cdot\left(\;-\;\efbold\;\times\;
 \frac{\partial {\efbold}}{\partial \Omega}\;
 \right)\;\ket\;=  \\
 \nonumber\\
 \nonumber\\
 \nonumber
 \frac{a^2}{4}\;\left\{\;
 \dot{\mu}_1\;\sin\inc\;
\left[\;-\;\left(2\,+\,3\,e^2\right)\;\sin\Omega\;\cos\inc\;+\;5\;e^2\;\left(\,
\cos \Omega\;\sin 2\omega\,+\,\sin\Omega\;\cos 2\omega\;\cos
\inc \,\right) \;\right] \right. \\
 \nonumber\\
 \nonumber\\
 \nonumber
 +\;\dot{\mu}_2\;\sin\inc\;\left[\;
 \left(2\,+\,3\,e^2\right)\;\cos\Omega\;\cos\inc\;+\;5\,e^2\;\left(
 \sin\Omega\;\sin 2\omega\;-\;\cos\Omega\;\cos 2\omega\;\cos\inc
 \right)
 \;\right]~~~~\\
 \nonumber\\
 \nonumber\\
 \left. \;-\;\dot{\mu}_3\;\left[\;
 \left(2\,+\,3\,e^2\right)\,\left(2\;-\;\sin^2\inc\;\right)\;+\;
 {5}\,e^2\,\sin^2\inc\;\cos 2\omega
 \;\right]
 \;\right\}
 ~~~~~~~~~~~~~~~~~~~~~~~~~~~~~~~~~~~
 \label{900}
 \ea
Interestingly, even in the limit of vanishing eccentricity this
sum survives and becomes
 \ba
 \frac{a^2}{2}\,\left\{\,-\,
 \dot{\mu}_1\,\sin\inc\;\sin\Omega\;\cos\inc\,+\,
 \dot{\mu}_2\,\sin\inc\;\cos\Omega\;\cos\inc\,-\,
 \dot{\mu}_3\,\left(2\,-\,\sin^2\inc \right)
 \,\right\}\,=\,\frac{a^2}{2}\;\dot{\mu}_n\,\sin\inc\,-\,a^2\,\dot{\mu}_3\;
 \;.~~~
 \label{901}
 \ea
Moreover, even when both the eccentricity and inclination are nil,
this sum still remains nonzero and is equal to
$\;-\,\dot{\mu}_3\,a^2\,\;$.

 \ba
 \nonumber
 \mbox{{\bf{\large{\textbf{A.6.~~~Calculation of the secular and
 long-period parts of}}}}}
 \;\;\dotmubold\cdot\left(\;-\;\efbold\;\times\;
 \frac{\partial {\efbold}}{\partial \inc}\;
 \right)\;\;.~
 \ea

 \underline{The first term}:\\
 ~\\
 \ba
 %
 % THIS EXPRESSION HAS UNDERGONE A COMPREHENSIVE CHECK
 %
 \nonumber
 \bra\;\dot{\mu}_1\;\left(\;\frac{\partial \efbold}{\partial \inc}\;\times\;
 \efbold\;\right)_1\;\ket\;=\;~~~~~~~~~~~~~~~~~~~~~~~~~~~~~~~~~~~~~~~~~~
 ~~~~~~~~~~~~~~~~~~~~~~~~~~~~~~~~~~~~~~~~~~~\\
 \label{A87}\\
 \nonumber\\
 \nonumber
 \bra\;\dot{\mu}_1\;\,a^2\;\;\frac{\left(1\;-\;e^2\right)^2}{\left(1\;+\;e\;\cos
\nu\right)^2}\;\;\left[\;
           -\;\cos\Omega\;\sin(\omega +\nu)\;-\;
           \sin \Omega\;\cos(\omega +\nu)\;\cos \inc
  \;\right]\;\sin(\omega + \nu)\;\ket\;=\;\;\;\;
 \ea
 ~\\
 \ba
 \nonumber
 \dot{\mu}_1\;\,a^2\;\;\left(1\;-\;e^2\right)^2\;\;
 \frac{\left(1\;-\;e^2\right)^{3/2}}{2\;\pi}\;\;\left[\;
 -\;\Upsilon_2\;\cos\Omega\;-\;\Upsilon_1\;\sin\Omega\;\cos\inc \;\right]\;=
 ~~~~~~~~~~~~~~~~~~~~~~~~~~~~~~~~~~\\
 \nonumber\\
 \nonumber\\
 \nonumber
 \dot{\mu}_1\;a^2\;\;
 \frac{\left(1-e^2\right)^{7/2}}{2\;\pi}\;\left\{\;
 -\;\left[\,\frac{\pi}{2}\;\;\frac{2+3e^2-5e^2\;\cos
 (2\omega)}{\left(1-e^2\right)^{7/2}}\;\right]
 \,\cos\Omega\;-\;\left[\,\frac{5}{2}\;\pi\,e^2\;\frac{\sin(2\omega)}{
 \left(1-e^2\right)^{7/2}} \,\right]\;\sin\Omega\;\cos\inc
 \;\right\}\\
 \nonumber\\
 \nonumber\\
 \nonumber
 =\;\dot{\mu}_1\;\;\frac{a^2}{4}\;\left\{\;-\;\left(2\,+\,3\,e^2\right)
 \,\cos\Omega\,+\,5\,e^2\;\left[\;\cos\Omega\;\cos(2\omega)\,-\,\sin
 \Omega\;\sin(2\omega)\;\cos\inc \;\right]
 \;\right\}
 \ea
 ~\\
 ~\\
 \underline{The second term:}\\
 ~\\
 \ba
 %
 % THIS EXPRESSION HAS UNDERGONE A COMPREHENSIVE CHECK
 %
 \nonumber
 \bra\;\dot{\mu}_2\;\left(\;\frac{\partial \efbold}{\partial \inc}\;\times\;
 \efbold\;\right)_2\;\ket\;=~~~~~~~~~~~~~~~~~~~~~~~~~~~~~~~~~~~~~~~~~~~~~~~~~~~~
 ~~~~~~~~~~~~~~~~~~~~~~~~~~~~~~~~~\\
 \label{A88}\\
 \nonumber\\
 \nonumber
 \bra\;\dot{\mu}_2\;\,a^2\;\;\frac{\left(1\;-\;e^2\right)^2}{\left(1\;+\;e\;\cos
 \nu\right)^2}\;\; \left[\;-\;\sin\Omega\;\sin(\omega +\nu)\;+\;
                    \cos \Omega\;\cos(\omega +\nu)\;\cos \inc
                   \;\right]\;\sin (\omega + \nu)\;\ket\;=\;\;\;
 \ea
 ~\\
 \ba
 \nonumber
 \dot{\mu}_2\;\,a^2\;\;\left(1\;-\;e^2\right)^2\;\;
 \frac{\left(1\;-\;e^2\right)^{3/2}}{2\;\pi}\;\;\left[\;
 -\;\Upsilon_2\;\sin\Omega\;+\;\Upsilon_1\;\cos\Omega\;\cos\inc \;\right]\;=
 ~~~~~~~~~~~~~~~~~~~~~~~~~~~~~~~~~~\\
 \nonumber\\
 \nonumber\\
 \nonumber
 \dot{\mu}_2\;a^2\;\;
 \frac{\left(1-e^2\right)^{7/2}}{2\;\pi}\;\left\{\;
 -\;\left[\,\frac{\pi}{2}\;\;\frac{2+3e^2-5e^2\;\cos
 (2\omega)}{\left(1-e^2\right)^{7/2}}\;\right]
 \,\sin\Omega\;+\;\left[\,\frac{5}{2}\;\pi\,e^2\;\frac{\sin(2\omega)}{
 \left(1-e^2\right)^{7/2}} \,\right]\;\cos\Omega\;\cos\inc
 \;\right\}\\
 \nonumber\\
 \nonumber\\
 \nonumber
 =\;\dot{\mu}_2\;\;\frac{a^2}{4}\;\left\{\;-\;\left(2\,+\,3\,e^2\right)
 \,\sin\Omega\,+\,5\,e^2\;\left[\;\sin\Omega\;\cos(2\omega)\,+\,\cos
 \Omega\;\sin(2\omega)\;\cos\inc\; \right]
 \;\right\}
 \ea
% ~\\
 ~\\
 \underline{The third term:}\\
~\\
 \ba
 %
 % THIS EXPRESSION HAS UNDERGONE A COMPREHENSIVE CHECK
 %
 \bra\;\dot{\mu}_3\;\left(\;\frac{\partial \efbold}{\partial \inc}\;\times\;
 \efbold\;\right)_3\;\ket\;=\;\bra\;\dot{\mu}_3\;\,a^2\;
 \frac{\left(1 - e^2\right)^2}{\left(1 + e\;\cos \nu\right)^2}\;\,
 \sin(\omega + \nu)\;\cos(\omega + \nu)\;\sin\inc\;\ket ~=~~~~~~~~~~~~~~~~~~~~
 \label{A89}
 \ea
 ~\\
 \ba
 \nonumber
 \dot{\mu}_3\;\,a^2\;\;\left(1\;-\;e^2\right)^2\;\;
 \frac{\left(1\;-\;e^2\right)^{3/2}}{2\;\pi}\; \Upsilon_1\;\sin\inc\;=\;
 \dot{\mu}_3\;a^2\;\;
 \frac{\left(1-e^2\right)^{7/2}}{2\;\pi}\;
 \,\frac{5}{2}\;\pi\,e^2\;\frac{\sin(2\omega)}{
 \left(1-e^2\right)^{7/2}}\;\sin\inc\; \\
 \nonumber\\
 \nonumber\\
 \nonumber
 =\;\dot{\mu}_3\;a^2\;\frac{5}{4}\;e^2\;\sin 2\omega\;\sin\inc
 ~~~~~~~~~~~~~~~~~~~~~~~~~~~~~~~~~~~~~~~~~~~~~~~~~~~~~~~~~~~~~~~~~~~~~
 ~~~~~~~~~~~~~~~~~~
 \ea
 ~\\
% ~\\
 \underline{Total:}
 \ba
 \nonumber
 \bra\;\dotmubold\cdot\left(\;-\;\efbold\;\times\;
 \frac{\partial {\efbold}}{\partial \inc}\;
 \right)\;\ket\;=~~~~~~~~~~~~~~  \\
 \label{A90}\\
 \nonumber\\
 \nonumber
 \frac{a^2}{4}\;\left\{\;
 \dot{\mu}_1\;\left[\;-\;\left(2\,+\,3\,e^2\right)
 \,\cos\Omega\,+\,5\,e^2\;\left(\cos\Omega\;\cos 2\omega\,-\,\sin
 \Omega\;\sin 2\omega\;\cos\inc \right)
 \;\right]\;+ \right.\\
 \nonumber\\
 \nonumber\\
 \nonumber
 \dot{\mu}_2\;\left[\;-\;\left(2\,+\,3\,e^2\right)
 \,\sin\Omega\,+\,5\,e^2\;\left(\sin\Omega\;\cos 2\omega\,+\,\cos
 \Omega\;\sin 2\omega\;\cos\inc \right)
 \;\right]\;+\\
 \nonumber\\
 \nonumber\\
 \nonumber
 \left. \dot{\mu}_3\;\left[\;5\;e^2\;\sin
 2\omega\;\sin\inc
 \;\right]\;\right\}~~~~~~~~~~~~~~~~~~~~~~~~~~~~~~~~~~~~~~~~~~~~~~~~~~~~~~~~~~
 \ea
 In the limit of vanishing eccentricity this sum approaches
 \ba
 -\;\frac{a^2}{2}\;\left\{\;\dot{\mu}_1\;\cos\Omega\;+
 \;\dot{\mu}_2\;\sin\Omega  \;\right\}\;\;\;,
 \label{903}
 \ea
expression that bears no dependence upon the inclination.

 \ba
 \nonumber
 \mbox{{\bf{\large{\textbf{A.7.~~~Calculation of the secular part of}}}}}
 \;\;\;\;\dotmubold\cdot\left(\;-\;\efbold\;\times\;
 \frac{\partial {\efbold}}{\partial M_o}\;
 \right)\;\;\;.~~~~~~~~~~~~~~~~~
 \ea

The considered quantity, when expressed through the orbital
elements, looks as
 \ba
 \dotmubold\;\cdot\;\left(\;-\;\efbold\;\times\;
 \frac{\partial {\efbold}}{\partial
 M_o}\;\right)\;=\;
 -\;\dot{\mu}_{\perp}\;a^2\;\;\sqrt{1\,-\,e^2}\;\;\;.\;\;\;\;\;
 \label{A91}
 \ea
As we are trying to calculate not the exact values of the elements
but their secular parts, we assume that, at shorter than an
orbital period time scales, $\,a\,,\;e\,$ and
$\,\dot{\mu}_{\perp}\,$ stay unchanged. In this approximation, the
above expression coincides with its secular part:
 \ba
 \bra\;\dotmubold\;\cdot\;
 \left(\;-\;\efbold\;\times\;\frac{\partial {\efbold}}{\partial
 M_o}\;\right)\;\ket\;=\;
 -\;\dot{\mu}_{\perp}\;a^2\;\;\sqrt{1\,-\,e^2}\;\;\;.
 \label{A92}
 \ea
~\\
~\\
 \ba
 \nonumber
 \mbox{{\bf{\large{\textbf{A.8.~~~The planetary precession rate
  $ \;\mubold\; $ and its projection  $ \;{{\mu}}_{\perp}\; $}}}}}\\
  \nonumber
  \mbox{{\bf{\large{onto the satellite's orbital momentum.
  ~~~~~~~~~~~~~~~~~~~~~}}}}
  \ea\\

 Let the inertial axes $\;(\,X\,,\;Y\,,\;Z\,)\;$ and the corresponding
 unit vectors $\;(\,\mathbf{\hat{X}}\,,\;\mathbf{\hat{Y}}\,,\;
 \mathbf{\hat{Z}}\,)
 \;$ be fixed in space so that $\;X\;$ and $\;Y\;$ belong to the equator of
 epoch.
 A rotation within the equator-of-epoch plane by longitude $\;h_p\;$, from
 axis $\;X\;$, will define the line of nodes, $\;x\;$. A rotation about
 this line by an inclination angle $\;I_p\;$ will give us the planetary equator
 of date.
 The line of nodes $\;x\;$, along with axis $\;y\;$ naturally chosen within
 the
 equator-of-date plane, and with axis $\;z\;$ orthogonal to
 this plane, will constitute the precessing coordinate system, with the
 appropriate basis denoted by $\;(\,\mathbf{\hat{x}}\,,\;\mathbf{\hat{y}}\,,
 \;\mathbf{\hat{z}}\,)\;$.

 In the inertial basis $\;(\,\mathbf{\hat{X}}\,,\;\mathbf{\hat{Y}}\,,\;
 \mathbf{\hat{Z}}\,)
 \;$, the direction to the North Pole of date is given by
  \ba
  {\bf{\hat{z}}}\;=\;\left(\;\sin I_p\,\sin h_p\;\;,\;\;\;-\,\sin I_p\,\cos h_p
  \;\;,\;\,\;\cos I_p\;\right)^{^T}
  \label{z}
  \ea
 while the total angular velocity reads:
 \ba
 \omegabold^{(inertial)}_{total}\;=\;{\bf{\hat{z}}}\;s\;+\;
 \mubold^{(inertial)}
 \;\;\;,
 \label{}
 \ea
 the first term denoting the rotation about the precessing axis $\;{\bf{\hat{z}}}\;$, the second
 term being the precession rate of $\;{\bf{\hat{z}}}\;$ relative to the inertial frame $\;(\,
 \mathbf{\hat{X}}\,,\;\mathbf{\hat{Y}}\,,\;\mathbf{\hat{Z}}\,)\;$, and $\;s\;$ standing for the
 angular velocity of rotation about the axis $\;{\bf{\hat{z}}}\;$. This precession rate is given by
 \ba
 \mubold^{(inertial)}\;=\;\left(\;
 {\dot{I}}_p\,\cos h_p\;\;\;,\;\;\;\,{\dot{I}}_p\,\sin h_p
  \;\;\;,\;\;\,\;{\dot{h}}_p\;\right)^{^T}\;\;\;,
 \label{muinertial}
 \ea
 because this expression satisfies $\;\mubold^{(inertial)}\;\times\;{\bf{\hat{z}}} \;=\;
 {\bf{\dot{\hat{z}}}}\;$.

 In a frame co-precessing with the equator of date, the precession
 rate will be represented by vector
 \ba
 \mubold\;=\;{\bf{\hat{R}}}_{i\rightarrow p}\;\;\mubold^{(inertial)}\;\;\;,
 \label{}
 \ea
 where the matrix of rotation from the equator of epoch to that of date (i.e.,
 from the inertial frame to the precessing one) is given by
   \ba
  {\bf{\hat{R}}}_{i\rightarrow p}\;=\;
 \left[
 \begin{array}{ccc}
    ~~~~~~~~~~~~~~~~\, \cos h_p   &    ~~~~~~~~~~~~~~~~\,\sin h_p          &  ~~~~~~~~~~~~ 0 \\
\\
  ~~~\, -\;\cos I_p \;\sin h_p &  \;\;\;\;\;\;\;\;\;\,\cos I_p\;\,\cos h_p  &   ~~~~~~~~~\sin I_p  \\
\\
  ~~~~~~~~~\;\,\sin I_p\;\,\sin h_p\;\;\;  & \;\;\;\;\;-\;\sin I_p\,\;\cos h_p  &  ~~~~~~~~~\cos I_p
\end{array}
 \right]
 \label{}
 \ea
 From here we get the components of the precession rate, as seen
 in the co-precessing coordinate frame $\;(x\,,\;y\,,\;z)\;$:
 \ba
 \mubold\;=\;\left(\;\mu_1\;,\;\;\mu_2\;,\;\;\mu_3\;\right)^{^T}\;=\;\left(\;
 {\dot{I}}_p\;\;,\;\;\;\,{\dot{h}}_p\,\sin I_p
  \;\;,\,\;\;\;{\dot{h}}_p\,\cos I_p\;\right)^{^T}\;\;\;.
 \label{}
 \ea
 In our paper we also need the components of $\;\dotmubold\;$ dot standing for derivatives
 calculated in the frame co-precessing with the equator:
 \ba
 \dotmubold\;=\;\left(\;{\dot{\mu}}_1\;,\;\;{\dot{\mu}}_2\;,\;\;
 {\dot{\mu}}_3\;\right)^{^T}\;=\;\left(\;
 {\ddot{I}}_p\;\;,\;\;\;\,{\ddot{h}}_p\,\sin
 I_p\,+\;\dot{h}_p\;\dot{I}_p\;\cos I_p\;
  \;\;,\,\;\;\;{\ddot{h}}_p\,\cos I_p\,-\;{\dot{h}}_p\;{\dot{I}}_p\;
  \sin I_p\right)^{^T}\;\;\;.\;\;\;
 \label{}
 \ea
% and
% \ba
% \nonumber
% {\bf{\ddot{\mubold}}}\,=\,\left(\,{\ddot{\mu}}_1\;,\;\,{\ddot{\mu}}_2\;,\;\,{\ddot{\mu}}_3
% \,\right)^{^T}\,=\,\left({\stackrel{...}{\textstyle{I}}}_p\,,\;\;{\stackrel{...}{
% \textstyle{h}}}_p\,\sin I_p + {\ddot{h}}_p {\dot{I}}_p\,\cos I_p\,+\,\ddot{h}_p\dot{I}_p\,
% \cos I_p +\dot{h}_p \ddot{I}_p\,\cos I_p -\dot{h}_p \dot{I}_p \dot{I}_p\,\sin I_p\,, \right.
% \\  \label{}\\
% \nonumber
% \left.
% {\stackrel{...}{\textstyle{h}}}_p\,\cos I_p-{\ddot{h}}_p {\dot{I}}_p\,\sin I_p-\ddot{h}_p
% \dot{I}_p\,\sin I_p-\dot{h}_p \ddot{I}_p\,\sin I_p-\dot{h}_p \dot{I}_p\,\dot{I}_p\,\cos I_p
% \right)^{^T}~~~~~~~
% \label{}
% \ea
 The matrix of rotation from the precessing frame of the equator
 of date to the frame associated with the satellite's orbital
 plane will look:
  \ba
 {\bf{\hat{R}}}_{p\rightarrow o}\;=\;
 \left[
 \begin{array}{ccc}
    ~~~~~~~~~~~~~~~~ \cos \Omega   &    ~~~~~~~~~~~~~~~~~~\sin \Omega  &  ~~~~~~~~~~~ 0 \\
 \\
  ~~~\,-~\cos \inc ~\sin \Omega & ~~~~~~~~~\,\cos \inc~\,\cos \Omega & ~~~~~~~~~\sin {\inc}  \\
 \\
  ~~~~~~~~~~\,\sin \inc~\,\sin\Omega ~~~ & ~~~~~-~\sin \inc\,~\cos \Omega & ~~~~~~~~~\cos \inc
 \end{array}
 \right]
 \label{A19}
 \ea
 This will give us the precession rate as seen in the instantaneous orbit frame:
 \ba
 \mubold^{(orb)}\;=\;{\bf{\hat{R}}}_{p\rightarrow
 o}\;\;\mubold\;\;\;.
 \label{}
 \ea
 The second component of this vector (i.e., the component pointing
 toward the ascending node of the satellite orbit relative to the
 equator of date) is what we need in our formulae (\ref{52}) and
 (\ref{54}):
 \ba
 \nonumber
 \mu_n\;=\;\;-\;{\mu}_1\;\sin \Omega\;\cos\inc\,+\,
 {\mu}_2\;\cos \Omega\;\cos \inc\,+\,{\mu}_3\;\sin
 \inc~~~~~~~~~~~~~~~~~~ \\
 \label{}\\
 \nonumber
 =\;-\;\dot{I}_p\;\sin \Omega\;\cos\inc\,+\,
 \dot{h}_p\,\sin I_p\;\cos \Omega\;\cos \inc\,+\,
 \dot{h}_p\,\cos I_p\;\sin\inc\;\;\;.
 \ea
 Its time derivative (taken in the frame of reference co-precessing with the
 equator of date) is:
  \ba
  \nonumber
 {\dot{\mu}}_n\;=\;-\;\dot{\mu}_1\;\sin \Omega\;\cos\inc\,+\,\dot{\mu}_2\;
 \cos\Omega\;\cos \inc\,+\,\dot{\mu}_3\;\sin \inc~~~~~~~~~~~~~~~~~~~~~~~~~~~~~~~~
 ~~~~~~~~~~~~~~~~~~~~~~~~~~~~ \\
 \label{}\\
 \nonumber
 =\;-\;\ddot{I}_p\;\sin \Omega\;\cos\inc\,+\,
 \left(\,
 {\ddot{h}}_p\,\sin I_p\,+\,{\dot{h}}_p\,{\dot{I}}_p\,\cos I_p
 \,\right)
 \;\cos \Omega\;\cos \inc\,+\,
 \left(\,
 {\ddot{h}}_p\,\cos I_p\,-\,{\dot{h}}_p\,{\dot{I}}_p\,\sin I_p
 \,\right)
 \;\sin
 \inc\;\;.\;\;
 \label{}
 \ea
 The third component of $\;\mubold^{(orb)}\;$ (i.e., the component orthogonal to
 the instantaneous plane of the orbit) is exactly what we need in (\ref{27})
 and (\ref{33}):
 \ba
 \nonumber
 \mu_{\perp}\;=\;\mu_1\;\sin \inc\;\sin \Omega\;-
 \;\mu_2\;\sin \inc\;\cos \Omega\;+\;\mu_3\;\cos \inc ~~~~~~~~~~~~~~~~\\
 \label{}\\
 \nonumber
 =\;{\dot{I}}_p\;\sin \inc\;\sin \Omega\;-
 \;{\dot{h}}_p\;\sin I_p\;\sin \inc\;\cos \Omega\;+
 \;{\dot{h}}_p\;\cos I_p\;\cos \inc
 \ea
 Its time derivative $\;{\dot{\mu}}_{\perp}\;$ defined in
 the axes co-precessing with the equator (and therefore equal to
 $\;\dot{\mubold}\cdot\wbold\;$, not to $\;d(\mubold\cdot\wbold)/dt\;$)
 will now be expressed by
 \ba
 \nonumber
 {\dot{\mu}}_{\perp}\;\equiv\;\dot{\mu}_1\;\sin \inc\;\sin \Omega\;-
 \;\dot{\mu}_2\;\sin \inc\;\cos \Omega\;+\;\dot{\mu}_3\;\cos \inc\;=
 ~~~~~~~~~~~~~~~~~~
 \\
 \nonumber\\
 {\ddot{I}}_p\,\sin \inc\;\sin \Omega\,-
 \,\left(\,
 {\ddot{h}}_p\,\sin I_p\,+\,{\dot{h}}_p\,{\dot{I}}_p\,\cos I_p
 \,\right)\,\sin \inc\,\cos \Omega\,+
 \,\left(\,
 {\ddot{h}}_p\,\cos I_p\,-\,{\dot{h}}_p\,{\dot{I}}_p\,\sin I_p
 \,\right)\,\cos \inc\;\;\;\;\;\;\;\;\;\;\;\;\;\;
 \label{}\\
 \nonumber\\
 \approx\;{\ddot{h}}_p\,\left(\,
 -\,\sin I_p\;\sin\inc\;\cos\Omega\;+\;\cos I_p\;\cos\inc
 \,\right)~~~,~~~~~~~~~~~~~~~~~~~~~~~
 \label{}
 \ea
% while $\;\ddot{\mu}_{\perp}\;$ will look as
% \ba
% \nonumber
% {\ddot{\mu}}_{\perp}\;\equiv\;\ddot{\mu}_1\;\sin \inc\;\sin \Omega\;-
% \;\ddot{\mu}_2\;\sin \inc\;\cos \Omega\;+\;\ddot{\mu}_3\;\cos \inc\;=~~~~~~~~~~~~~~~~~~
% \\
% \nonumber\\
% \nonumber
% {\stackrel{...}{\textstyle{I}}}_p\;\sin \inc\;\sin \Omega\;-\;\left(\;
% {\stackrel{...}{\textstyle{h}}}_p\,\sin I_p + {\ddot{h}}_p {\dot{I}}_p\,\cos I_p\,+\,
% \ddot{h}_p \dot{I}_p\,\cos I_p +\dot{h}_p \ddot{I}_p\,\cos I_p -
% \dot{h}_p \dot{I}_p \dot{I}_p\,\sin I_p\;\right)\;\sin \inc\;\cos \Omega\\
% \nonumber\\
% +\;\left(\;{\stackrel{...}{\textstyle{h}}}_p\,\cos I_p - {\ddot{h}}_p {\dot{I}}_p\,
% \sin I_p -\ddot{h}_p \dot{I}_p\,\sin I_p -\dot{h}_p \ddot{I}_p\,\sin I_p -
% \dot{h}_p \dot{I}_p\,\dot{I}_p\,\cos I_p\;\right)\;\cos \inc\;\;\;\;\;\;\;\;
% \label{}\\
% \nonumber\\
% \approx\;{\stackrel{...}{\textstyle{h}}}_p\;\left(\,-\,\sin I_p\;\sin\inc\;\cos\Omega
% \;+\;\cos I_p\;\cos\inc\,\right)~~~,~~~~~~~~~~~~~~~~~~~~~~~
% \label{}
% \ea
 $I_p\;$ and $\;h_p\;$ being the inclination and the longitude of
 the node of the equator of date relative to the one of epoch.

 The expression for $\;\dot{\mu}_{\perp}\;$  permitted
 approximations shown above because, for Mars' equator, the speed of the nodes'
 motion, $~|\dot{h}_p|~\approx~360^o/(1.75\times 10^5\,yr)~\approx\;2\times 10^{
 -3}\;yr^{-1}\;$, much exceeds the rate of its inclination change, $~|\dot{I}_p|~
 \approx\;5^o/(0.5\times 10^6\,yr)\;\approx\;10^{-5}\;yr^{-1}\;$ (Ward 1974).
~\\
~\\
 \ba
 \nonumber
 \mbox{{\bf{\large{ {\textbf{A.9.~~~Calculation of}} $\;h_p\;$ {\textbf{and}}
 $\;I_p\;$.~~~~~~~~~~~~~~~~~~~~~~~~~~~~~~~~~~~~~~~~}}}}\\
  \nonumber
  \ea

The question now becomes as to how to calculate the time
dependence of $\;h_p\;$ and $\;I_p\;$. As very well known, these
two angles evolve in time because a non-spherical planet behaves
itself as an unsupported top whose precession is instigated by the
solar torque. The torques produced by the satellites are
irrelevant (Laskar 2004), the cases of the Moon and Charon being
exceptional. When the moments of inertia of the planet relate as
$\;C\;>\;B\;=\;A\;$, the solar torque is
 \ba
 {\bf\vec
 T}\;=\;\frac{3\;G\;M}{R^3}\;\,\left(C\;-\;A\right)\;
 \left({\bf\hat r}\cdot{\bf\hat k} \right)\;
 \left({\bf\hat r}\times{\bf\hat k} \right)\;\;\;,
 \label{}
 \ea
while in the general case of $\;C\;\geq\;B\;\geq\;A\;$ it is equal
to
 \ba
 {\bf\vec T}\;=\;\frac{3\;G\;M}{R^3}\;\,\left(C\;-\;\frac{A\,+\,B}{2}
 \right)\;
 \left({\bf\hat r}\cdot{\bf\hat k} \right)\;
 \left({\bf\hat r}\times{\bf\hat k} \right)\;\;\;,
 \label{}
 \ea
provided that the spin mode is not too deviant from the principal
one, and that this spin is much faster than the planet's orbital
revolution about the Sun.

In the above two formulae, $\;M\;$ is the solar mass, $\;R\;$
denotes the distance between the centres of masses of the planet
and the Sun, the unit vector $\;{\bf{\hat{r}}}\;$ points from the
planet toward the Sun, and the unit vector $\;{\bf{\hat{k}}}\;$
points in the direction of the major-inertia axis of the planet.

The precession of the angular momentum $\;\bf\vec L\;$ of the
planetary spin obeys
 \ba
 \frac{d{\bf\vec L}}{dt}\;=\;{\bf\vec T}
 \label{}
 \ea
 Colombo (1966) averaged this equation over the planet's year, under the assumption that the
 perturbing torque causes only very small variations of spin. This averaging yields the
 Colombo equation valid at timescales much exceeding one year:
 \ba
 \frac{d{\bf\vec L}}{dt}\;=\;
 \frac{3\;n^2_p}{2\;\left(1\,-\,e_p^2\right)^{3/2}}\;\,
 \left(C\;-\;\frac{A\,+\,B}{2} \right)\;
 \left({\bf\hat k}\cdot{\bf\hat n} \right)\;
 \left({\bf\hat k}\times{\bf\hat n} \right)\;\;\;,
 \label{Colombo_1}
 \ea
$\;n_p\;$ and $\;e_p\;$ being the mean motion and the eccentricity
of the planet's orbit about the Sun, and $\;{\bf\hat n} \;$ being
the unit vector normal to the planetary orbit; while $\;d{\bf\vec
L}/{dt}\;$ should be understood as a change of $\;{\bf\vec L}\;$
over a year, divided by the length of the year: $\;\Delta {\bf\vec
L}/{P}\;$. The angular velocity of the planet about its axis being
denoted with letter $\;s\;$, the Colombo equation may be rewritten
as
 \ba
 \frac{1}{s\;C}\;\frac{d{\bf\vec L}}{dt}\;=\;\alpha\;
 \left({\bf\hat k}\cdot{\bf\hat n} \right)\;
 \left({\bf\hat k}\times{\bf\hat n} \right)\;\;\;,
 \label{Colombo_2}
 \ea
the factor $\;\alpha\;$ being defined as
 \ba
 \alpha\;\equiv\;\frac{3\;n^2_p}{2\;s\;\left(1\,-\,e_p^2\right)^{3/2}}\;\,
 \frac{C\;-\;\frac{\textstyle A\,+\,B}{\textstyle 2} }{C}\;=\;
\frac{2\,\pi}{P}\;\,\frac{D}{P}\,\;\frac{1}{\left(1\,-\,e_p^2\right)^{3/2}}
 \left[\;\frac{3}{2}\;\;\frac{C\;-\;\frac{\textstyle A\,+\,B}{\textstyle 2} }{C}
 \;\right]
 \label{alpha_1}
 \ea
where $\;P\;=\;2\,\pi/n_p\;$  is the duration of the planet's
year, and $\;D\;=\;2\,\pi/s\;$ is that of its day. The relative
difference between the moments of inertia may be expressed through
the parameter $\;J_2\;$ emerging in the expression for potential
via the planetocentric latitude $\,\phi\,$
 \ba
 V\,=\;-\;\frac{G\,m}{r}\,
 \left[1\,-\,\sum_{n=2}^{\infty}\,J_n\,\left(\frac{\rho_e}{r}
 \right)^n\;P_n(\sin \phi)
 \right]\,+\,\sum_{n=2}^{\infty}\;\sum^{n}_{j=1}\,J_{nj}\,\left(\,
 \frac{\rho_e}{r}\,\right)^n\,P_{nj}\left(\sin \phi \right)\;\cos j\left(\lambda
 \,-\,\lambda_{nj} \right)
 \;\;,\;\;
 \label{}
 \ea
(where $\;\rho_e\;$ stands for the mean {\it{equatorial}} radius
of the planet):
 \ba
 J_2\;=\;\frac{C\;-\;\frac{\textstyle A\,+\,B}{\textstyle 2} }{M\;\rho_e^2}
 \;=\;\frac{C\;-\;\frac{\textstyle A\,+\,B}{\textstyle 2} }{C}\;
 \;\frac{C}{M\;\rho_e^2}\;\;\;.
 \label{}
 \ea
 It is also interconnected with the nonsphericity parameter
 $\;J\;$:
 \ba
 J\;\equiv\;\frac{3}{2}\;\left(\,\frac{\rho_e}{\rho}\,\right)^2 J_2\;=\;
 \frac{3}{2}\;\frac{C\;-\;\frac{\textstyle A\,+\,B}{\textstyle 2} }{C}\;
 \;\frac{C}{M\;\rho^2}\;=\;
 \frac{3}{2}\;\frac{C\;-\;\frac{\textstyle A\,+\,B}{\textstyle 2} }{C}\;
 \;{K}\;\;\;,
 \label{}
 \ea
 where $\;\rho\;$ is simply the mean (not the mean equatorial)
 radius of the planet, while the quantity
 \ba
 K\;\equiv\;\frac{C}{m\;\rho^2}
 \label{K}
 \ea
is the squared ratio of the gyration radius $\;\sqrt{C/m}\;$ of
the planet to its mean radius $\;\rho\;$. We thus see that
 \ba
 \frac{3}{2}\;\;\frac{C\;-\;\frac{\textstyle A\,+\,B}{\textstyle 2}}{C}\;
 =\;\frac{J}{K}\;=\;\frac{3}{2}\;{J_2}\;
 \frac{\,m\,\rho_e^2}{C}\,\;\;,
 \label{}
 \ea
 whence
 \ba
 \alpha\;=\;\frac{2\,\pi}{P}\;\,\frac{D}{P}\,\;
 \frac{1}{\left(1\,-\,e_p^2\right)^{3/2}}\;\,
 \frac{J}{K}\;=\;\frac{2\,\pi}{P}\;\,\frac{D}{P}\,\;
 \frac{1}{\left(1\,-\,e_p^2\right)^{3/2}}\;\,
 \frac{3}{2}\;{J_2}\;
 \frac{\,m\,\rho_e^2}{C}\,\;\;.
 \label{alpha_2}
 \ea
Ward (1974) used, for Mars, $\;K\;=\;0.359\;$ and
$\;J\;=\;2.95\,\times\,10^{-3}\;$, which gave him the
value:\footnote{~In his formulae, Ward (1973) missed or
deliberately neglected the factor
$\;\left(1\,-\,e_p^2\right)^{-3/2}\,$. In the case of Mars, this
may look legitimate because nowadays this factor amounts to
$\;1.013\,$. However, the Martian eccentricity is wont to have
varied through aeons within the interval of
$\;e\;=\;0.01\;\div\;0.14\;$ (Murray et al 1973). This means that
the said factor, $\;\left(1\,-\,e_p^2\right)^{-3/2}\,$, might have
varied from almost unity through $\;1.03\;$. This $\;3\,\%\;$
increase will look less than innocent if we recall that several
authors (Ward 1974, Laskar \& Robutel 1993, Touma \& Wisdom 1994)
insist on the stochastic nature of Mars' obliquity variations.}
$\;\alpha_{_{Mars}}\;=\;1.26\,\times\,10^{-12}\;rad/s\;=\;
8.19\;arcsec/yr\;$.

To further simplify the above expression (\ref{Colombo_2}),
Colombo (1966) assumed that the spin angular momentum $\;\bf\vec
L\;$ is parallel to the spin angular velocity $\;\bf\vec s\;$:
 \ba
 {\bf\vec L}\;\approx\;C\;{\bf\vec s}\;\approx\;C\;s\;{\bf\hat k}
 \;\;\;.
 \label{approx}
 \ea
While investigating the dynamics of the Moon at less than cosmic
time scales, Colombo certainly could afford this approximation.
Compare the latter with the exact expression for $\;\bf\vec L\;$
through $\;\bf\vec s\;$ and through the moments of inertia
$\;C\;\geq\;B\;\geq\;A\;$:
 \ba
% \nonumber
 {\bf\vec L}\,=\,{\bf\hat i}\,s_1\,A\,+\,{\bf\hat j}\,s_2\,B\,+\,
 {\bf\hat k}\,s_3\,C\,=
% \,{\bf\hat i}\,s_1\,(A\,-\,C)\,+\,
% {\bf\hat j}\,s_2\,(B\,-\,C)\,+\,
% \left({\bf\hat i}\,s_1\,+\,{\bf\hat j}\,s_2\,+\,{\bf\hat
% k}\,s_3\right)\,C\,=\\
% \nonumber\\
% \nonumber
% {\bf\hat i}\,s_1\,(A\,-\,C)\,+\,
% {\bf\hat j}\,s_2\,(B\,-\,C)\,+\,{\bf\vec
% s}\,C\,=\,
{\bf\hat i}\,s_1\,(A\,-\,C)\,+\,
 {\bf\hat j}\,s_2\,(B\,-\,C)\,+\,C\,s\,\left({\bf\hat p}\,-\,{\bf\hat k}\right)
 ~+~C\,s\,{\bf\hat k}~~~,~~
% \\
% \nonumber\\
% \approx~C\,s\,{\bf\vec
% k}~~~,~~~~~~~~~~~~~~~~~~~~~~~~~~~~~~~~~~~~~~~~~~~
 \label{}
 \ea
 ${\bf\hat s}\;\equiv\;\left(\,{\bf\hat i}\,s_1\,+\,{\bf\hat j}\,s_2\,+\,
 {\bf\hat k}\,s_3\,\right)\,s^{-1}\;\;$ being the instantaneous direction of the
 angular velocity of the planet's spin. We see that Colombo's approximation stems
 from the frivolous assertion that the planet always remains in a principal spin
 state. Indeed, insofar as $\;\bf\hat s\;$ coincides with $\;\bf\vec k\;$ the
 components $\;s_1\;$ and $\;s_2\;$ are nil, and (\ref{approx}) becomes
 exact.  Under such an assertion, the equation for unit vector aimed in the direction of the
major-inertia axis, $\bf\vec k\;$, assumes the form\footnote{~By
basing his research on the
 approximation (\ref{k}), Ward (1974) implicitly made the strong
 assumption of Mars always remaining in the principal spin state,
 polar wander and nutations and the Chandler wobble being
 neglected. By employing a Hamiltonian, that generates equation (\ref{k}),
 Laskar \& Robutel (1993) and Touma \& Wisdom (1994), too, rested their
 study on the same assumption.}
 \ba
 \frac{d{\bf\hat k}}{dt}\;=\;\alpha\;\left({\bf{\hat{n}}}\cdot{\bf{\hat{k}}}
 \right)\,\left({\bf{\hat{k}}}\times{\bf{\hat{n}}}  \right)
 \label{k}
 \ea
the unit normal to the planet's orbit, $\;\bf\vec n\;$, being
subject to variations described by the formulae and tables worked
out by Brouwer \& van Woerkom (1950).\footnote{~Brouwer \& van
Woerkom (1950) chose the ecliptic of 1950 as the reference plane.
Since our eventual goal is to simply estimate the range of
variations of $\;\inc\;$ and $\;\Omega\;$ over large time scales,
we can accept, without loss of generality, that at some distant
epoch the Martian equator coincided with that plane.} The
quantities $\;h_p\,,\;I_p\;$ and their time derivatives can be
calculated from integration of the above equation for $\;\bf\hat
k\;$. To this end, let us recall that our unit vector $\;{\bf\hat
k }\;$ coincides with the afore discussed unit vector $\;{\bf\hat
z}\;$ (see formula (\ref{z}) above). Therefore, in the frame of
the equator of epoch (which we assume, for convenience, to
coincide with the ecliptic of 1950), $\;{\bf\hat k}\;$ and
$\;d{\bf\hat k}/dt\;$ will read:
 \ba
  {\bf{\hat{k}}}\;=\;\left(\;\sin I_p\,\sin h_p\;\;,\;\;\;-\,\sin I_p\,\cos h_p
  \;\;,\;\,\;\cos I_p\;\right)^{^T}
 \label{}
 \ea
 \ba
  \frac{d\bf{\hat{k}}}{dt}=\left(
  {\dot{I}}_p\,\cos I_p\,\sin h_p\,+\,{\dot{h}}_p\,\sin I_p\,\cos h_p
  \;\,,\;\;\,
  -\,{\dot{I}}_p\,\cos I_p\,\cos h_p\,+\,{\dot{h}}_p\,\sin I_p\,\sin h_p
  \;\,,\;\,\;
  -\,{\dot{I}}_p\,\sin I_p\right)^{^T}\;,\;\;~
 \label{}
 \ea
while the components of $\;\bf\hat n\;$,
 \ba
 {\bf\hat n}\;=\;\left(\;\sin I_{orb}\,\sin
\Omega_{orb}\;\;,\;\;\;-\,\sin I_{orb}\,\cos \Omega_{orb}
  \;\;,\;\,\;\cos I_{orb}\;\right)^{^T}\;
 \label{}
 \ea
may be expressed through the auxiliary variables
 \ba
 q\;=\;\sin I_{orb}\;\sin \Omega_{orb}\;\;\;,\;\;\;\;\;
 p\;=\;\sin I_{orb}\;\cos \Omega_{orb}
 \label{}
 \ea
whose evolution will be found from
 \ba
 q\;=\;\sum_{j=1}^{\infty} \,N_j\,\sin \left(s_j't\,+\,\delta_j
 \right)\;\;\;,
 \label{}
 \ea
 \ba
 p\;=\;\sum_{j=1}^{\infty} \,N_j\,\cos \left(s_j't\,+\,\delta_j
\right)\;\;\;.
 \label{}
 \ea
Under the assumption that the orbital elements are defined
relative to the ecliptic plane of 1950,  Brouwer \& van Woerkom
(1950) calculated the values of the amplitudes, frequencies, and
phases used in the above formulae. Below follow the triples of
numbers:
 % \\
 % ~\\

   \pagebreak

 \noindent
 ==============================================\\
 $j\;\;\;\;\;\;\;\;\;\;\;~~~~~~~~~~~~~~~~~~~N_j
 \;\;\;\;\;\;\;\;\;\;\;\;\;\;~~~~~~~~~~~~~~~~~~~~~
 s_j'\;\;\;\;\;\;\;\;\;\;
 ~~~~~~~~~~~~~~~~~~~~~~~~\delta_j' $\\
 ${\left.\,\right.}^{\left.\,\right.}~
 ~~~~~~~~~~~~~~~~~~~~~~~~~~~~~~~~~~~~~~~~~~~~~~~~~~~~~~~~~
 (arc\,sec/yr)~~~~~~~~~~~~~~~~~~~~~~~(deg)$\\
 ==============================================\\
$ 1\;\;\;\;\;\;\;\;\;\;\;~~~~~~~~~~~~~~~0.0084889
 \;\;\;\;\;\;\;\;\;\;\;\;~~~~~~~~~~~-\,5.201537\;\;\;\;\;\;\;\;\;\;
 ~~~~~~~~~~~~\,~19.43255 $\\
 ~\\
$ 2\;\;\;\;\;\;\;\;\;\;\;~~~~~~~~~~~~~~~0.0080958
 \;\;\;\;\;\;\;\;\;\;\;\;~~~~~~~~~~~-\,6.570802\;\;\;\;\;\;\;\;\;\;
 ~~~~~~~~~~~~318.05685 $\\
 ~\\
$ 3\;\;\;\;\;\;\;\;\;\;\;~~~~~~~~~~~~~~~0.0244823
 \;\;\;\;\;\;\;\;\;\;\;\;~~~~~~~~~~~-\,18.743586\;\;\;\;\;\;\;\;\;\;
 ~~~~~~~~~~255.03057 $\\
 ~\\
$ 4\;\;\;\;\;\;\;\;\;\;\;~~~~~~~~~~~~~~~0.0045254
 \;\;\;\;\;\;\;\;\;\;\;\;~~~~~~~~~~~-\,17.633305\;\;\;\;\;\;\;\;\;\;
 ~~~~~~~~~~296.54103 $\\
 ~\\
$ 5\;\;\;\;\;\;\;\;\;\;\;~~~~~~~~~~~~~~~0.0275703
 \;\;\;\;\;\;\;\;\;\;\;\;~~~~~~~~~~~~+\,0.000004\;\;\;\;
 ~~~~~~~~~~~~~\,~~107.10201 $\\
 ~\\
$ 6\;\;\;\;\;\;\;\;\;\;\;~~~~~~~~~~~~~~~0.0028112
 \;\;\;\;\;\;\;\;\;\;\;\;~~~~~~~~~~~-\,25.733549\;\;\;\;\;\;\;\;\;\;
 ~~~~~~~~~~127.36669 $\\
 ~\\
$ 7\;\;\;\;\;\;\;\;\;\;\;~~~~~~~~~~~-\,0.0017308
 \;\;\;\;\;\;\;\;\;\;\;\;~~~~~~~~~~~-\,2.902663\;\;\;\;\;\;\;\;\;\;
 ~~~~~~~~~~~315.06348 $\\
 ~\\
$ 8\;\;\;\;\;\;\;\;\;\;\;~~~~~~~~~~~-\,0.0012969
 \;\;\;\;\;\;\;\;\;\;\;\;~~~~~~~~~~~-\,0.677522\;\;\;\;\;\;\;\;\;\;
 ~~~~~~~~~~~202.29272 $\\
 ==============================================
 ~\\
% In this table we provide the values of $\,\delta'_j\,$, which
% differ from those given by Brouwer \& Clemence (1950) by a
% constant shift. This shift corresponds to a different choice of
% the coordinate axes within the ecliptic plane of 1950. We follow
% the axes choice offered by Ward (1994).

Technically, the computation of time evolution of $\;I_p\;$ and
$\;h_p\;$ can be implemented through a set of differential
equations obtained by substitution of  \ba
 {\bf\hat n}\,=\,\left(\;q\,,\;-\;p\,,\;u\;\right)^{^T}\;\;,\;\;\;\;q\;\equiv\;
 \sin I_{orb}\;\sin \Omega_{orb}\;\;,\;\;\;\;
 p\;\equiv\;\sin I_{orb}\;\cos
 \Omega_{orb}\;\;,\;\;\;\;u\;\equiv\;\cos I_{orb}~~,~~~~
 \label{odin}
 \ea
 \ba
 {\bf\hat k}\,=\,\left(\;Q\,,\;-\;P\,,\;U\;\right)^{^T}\;\;\;,\;\;\;\;\;Q\;=\;
 \sin I_{p}\;\sin h_{p}\;\;\;,\;\;\;\;\;
 P\;=\;\sin I_{p}\;\cos h_{p}\;\;,\;\;\;\;U\;\equiv\;\cos I_{p}~~,~~~~
 \label{dva}
 \ea
into the Colombo equation (\ref{k}). Here follow these equations:
 \ba
 \frac{dQ(t)}{dt}\;=\;-\;\alpha\;\left[\;q(t)\;Q(t)\,+\,p(t)\;P(t)\,+\,u(t)\;U(t)\;\right]\;
 \left[\;-\;p(t)\;U(t)\;+\;u(t)\;P(t)\; \right]\;\;\;,\;\;\;
 \label{}
 ~\\
 \frac{dP(t)}{dt}\;=\;\,\alpha\;\,
 \left[\;q(t)\,Q(t)\,+\,p(t)\,P(t)\,+\,u(t)\,U(t)\;\right]\;\,
 \left[\;u(t)\;Q(t)\;-\;q(t)\;U(t)\;\right]\;\;\;\;,\;\;\;\;\;\;\;\;\;\,
 \label{}
 ~\\
 \frac{dU(t)}{dt}\;=\;-\;\alpha\;\left[\;q(t)\;Q(t)\,+\,p(t)\,P(t)\,+\,u(t)\,U(t)\;\right]\;\left[
 \;-\;q(t)\;P(t)\;+\;p(t)\;Q(t) \;\right]\;\;\;,\;\;\;\;\;
 \label{}
 \ea
 where, at each time step, the following values of $\;q(t)\,,\;p(t)\,,$ and $\;u(t)\;$ are to
 be used:
 \ba
 q(t)\;=\;\sum_{j=1}^{\infty} \,N_j\,\sin \left(s_j't\,+\,\delta_j
 \right)\;\;\;,
 \label{}
 ~\\
 p(t)\;=\;\sum_{j=1}^{\infty} \,N_j\,\cos \left(s_j't\,+\,\delta_j
\right)\;\;\;,
 \label{}
 ~\\
 u(t)\;=\;\pm\;\sqrt{1\;-\;q(t)^2\;-\;p(t)^2}\;\;\;.
 \label{u}
 \ea
 The resulting values of $\;Q\,,\;P\,,\;U\;$, obtained through this integration,
 will, at each time step, give us the angles $\;I_p\;$
 and $\;h_p\;$ via formulae that follow from (\ref{odin} -
 \ref{dva}):
  \ba
 h_p\;=\;\arctan\frac{Q}{P}\;\;\;,\;\;\;\;I_p\;=\;\arccos U ~~~,~~~~
 \label{}
 \ea
 It is evident from (\ref{dva}) that the variables $\;Q(t)\,,\;P(t)\,,\;U(t)\;$
 obey the constraint
 \ba
 Q(t)^2\;+\;P(t)^2\;+\;U(t)^2\;=\;1\;~~;~~~~
 \label{}
 \ea
and therefore fulfilment of this constraint should be checked during
integration. Deviation from it will indicate accumulation of errors.
At each step, some attention will be needed also when the current
 value of $\;u(t)\;$ is evaluated. (We mean the choice of sign in (\ref{u}).)

Finally, it should be emphasised that the development by Brouwer \&
Clemence (1950) is limited in terms of precision and, therefore, in
terms of the time span over which it remains valid. A more accurate
and comprehensive development, with a validity span of tens of
millions of years, was recently offered by Laskar (1988). At the
future stages of our project, when developing a detailed physical
model of the satellite motion, we shall employ Laskar's results.

 % \pagebreak

{}

\end{document}